\documentclass[a4paper,11pt]{article}
\pdfoutput=1 % if your are submitting a pdflatex (i.e. if you have
\usepackage{jheppub} % for details on the use of the package, please
\usepackage[T1]{fontenc} % if needed
\usepackage{slashed,bbm,here}
\title{\boldmath Marginal deformations of 3d supersymmetric U($N$) model and broken higher spin symmetry}

\author[a]{Yasuaki Hikida}
\author[b]{Taiki Wada}

\affiliation[a]{Center for Gravitational Physics, Yukawa Institute for Theoretical Physics, Kyoto University,\\ Kyoto 606-8502, Japan}

\affiliation[b]{Department of Physical Sciences, College of Science and Engineering, Ritsumeikan University,\\Shiga 525-8577, Japan}

\emailAdd{yhikida@yukawa.kyoto-u.ac.jp}
\emailAdd{rp0017xp@ed.ritsumei.ac.jp}
\abstract{We examine the marginal deformations of double-trace type in 3d supersymmetric U$(N)$ model with $N$ complex free bosons and fermions. We compute the anomalous dimensions of higher spin currents to the $1/N$ order but to all orders in the deformation parameters by mainly applying the conformal perturbation theory. The 3d field theory is supposed to be dual to 4d supersymmetric Vasiliev theory, and the marginal deformations are argued to correspond to modifying boundary conditions for bulk scalars and fermions. Thus the modification should break higher spin gauge symmetry and generate the masses of higher spin fields. We provide supports for the dual interpretation by relating bulk computation in terms of Witten diagrams to boundary one in conformal perturbation theory.
	
}

\arxivnumber{1701.03563}
\preprint{YITP-17-3}

\begin{document}
	\maketitle
	\flushbottom

\section{Introduction}

Conformal field theories (CFTs) with large $N$ structure are quite important since they can be utilized to investigate holographic dual gravity theory among others. 
In particular, CFTs with higher spin symmetry broken in $1/N$ are useful to examine symmetry breaking in dual higher spin gauge theories.
In \cite{Hikida:2016wqj,Hikida:2016cla},  the critical O$(N)$ scalar model and the Gross-Neveu model \cite{Gross:1974jv,Moshe:2003xn} in $d$ dimensions were studied, and the anomalous dimensions of higher spin currents in these models were reproduced by applying the method of conformal perturbation theory. 
A famous example of higher spin gauge theory on AdS space is given by Vasiliev theory \cite{Vasiliev:1995dn,Vasiliev:1999ba,Vasiliev:2003ev}, and  the two models are supposed to be holographic dual to the type A and type B Vasiliev theories \cite{Klebanov:2002ja,Leigh:2003gk,Sezgin:2003pt}.
The anomalous dimensions of higher spin currents correspond to the masses of higher spin fields, and the CFT computation in the conformal perturbation theory was interpreted from the bulk theory in terms of Witten diagrams. 
This paper addresses a continuation of these works.

Higher spin gauge theory is expected to describe superstring theory in the tensionless limit \cite{Gross:1988ue}, and turning on string tension should correspond to breaking higher spin gauge symmetry.
In \cite{Hikida:2016wqj,Hikida:2016cla}, the double-trace type deformations of free O($N$) bosons and free U($N$) fermions were examined, but there are only two fixed points, where one of them is free and the other is interacting with higher spin symmetry broken in $1/N$.
Since the string tension is a parameter of theory, it should be mapped to a marginal deformation of CFT dual to higher spin gauge theory. 
Therefore it is better to work on a model which has higher spin dual and admits marginal deformations.
As a simple example, we examine $\mathcal{N}=2$ supersymmetic U$(N)$ model with $N$ free complex scalars and fermions in three dimensions.%
\footnote{Another type of examples are given by Chern-Simons-matter theories in three dimensions \cite{Giombi:2011kc,Aharony:2011jz,Maldacena:2012sf}, and the anomalous dimensions of higher spin currents have been computed in \cite{Giombi:2016zwa} recently.}
The model is proposed to be dual to a supersymmetric Vasiliev theory on AdS$_4$ \cite{Leigh:2003gk,Sezgin:2003pt}, and it admits deformations exactly marginal at least to the $1/N$ order.

We compute the anomalous dimensions of higher spin currents in the deformations of supersymmetric model to the $1/N$ order but to all orders in deformation parameters mainly with the help of conformal perturbation theory.
Furthermore, we interpret the CFT computation in terms of bulk higher spin theory.
Let us denote $\phi^i$ and $\psi^i$ $(i=1,2,\ldots,N)$ as the free complex bosons and fermions in the supersymmetric model. A marginal deformation of the theory is given by
\begin{align}
\Delta_\lambda S = \lambda \int d^3 x \mathcal{O} (x) \tilde{\mathcal{O}} (x)
\label{def}
\end{align}
with $\mathcal{O} = \bar \phi^i \phi_i$ and $\tilde{\mathcal{O}} = \bar \psi^i \psi_i$.
The scalar operators are dual to bulk scalar fields, and the deformation corresponds to mixing the boundary conditions of these scalars \cite{Witten:2001ua}. 
Higher spin symmetry is broken with non-zero $\lambda$, and the anomalous dimensions of higher spin currents are obtained to the $1/N$ order but to all orders in $\lambda$, see \eqref{anomalousb} below.
There is another type of marginal deformation as
\begin{align}
\Delta_{\kappa} S = \kappa \int d^3 x \bar{\mathcal{K}} (x) \mathcal{K} (x) 
\label{defsusy}
\end{align}
with $\mathcal{K} = \bar{\phi}^{i} \psi_i$ and $\bar{\mathcal{K}} = \bar \psi^i \phi_i$.
The fermionic operators are dual to bulk fermionic fields, and the deformation corresponds to mixing the boundary conditions of these spinors.
The anomalous dimensions of higher spin currents are computed to the $1/N$ order but to all orders in $\kappa$, see \eqref{anomalousf} below.%
\footnote{The anomalous dimensions are computed in conformal perturbation theory up to the $\kappa^2$ order. Higher order corrections in $\kappa$ are examined in a different method as used in \cite{Muta:1976js} for the Gross-Neveu model in $1/N$-expansion. The method is enough to obtain the all order expression of anomalous dimensions.
However, it is suitable to use the conformal perturbation theory if one wants to relate with dual higher spin theory. \label{muta}}
We can include both the deformations simultaneously, and supersymmetry is preserved when $2 \lambda = \kappa$.%
\footnote{The condition can be derived by writing the marginal deformation preserving supersymmetry with superfields and expanding it in terms of component fields as was done in \cite{Leigh:2003gk}. The same condition can be obtained also from the dual gravity theory. The boundary conditions of bulk scalars and spinors preserve supersymmetry when their deformation parameters are the same, see, e.g., \cite{Chang:2012kt}.  The parameters correspond to $\tilde \lambda,\tilde \kappa$ in \eqref{tlambda}, \eqref{tkappa}, thus the condition is deduced from $\tilde \lambda= \tilde \kappa$.}

\begin{figure}
	\centering
	\includegraphics[keepaspectratio, scale=0.5]
	{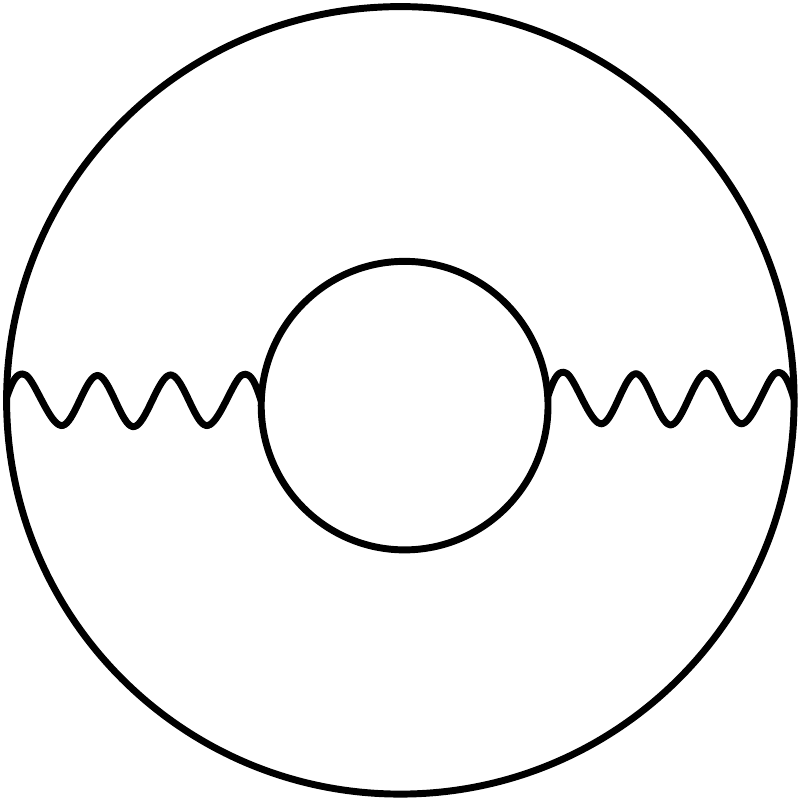}
	\caption{One-loop Witten diagram for the two point function of dual higher spin current.}
	\label{2pt}
\end{figure}

We adopt the conformal perturbation theory to compute the anomalous dimensions, which enables us to borrow the previous results in \cite{Hikida:2016wqj,Hikida:2016cla}.
The main reason to use the method is that the CFT computation can be understood in terms of bulk theory. The relation is known between the masses of higher spin fields and the anomalous dimensions of dual higher spin currents, which may be read off from two point function of the currents. The two point function is computable from the bulk theory through Witten diagrams and the contributions to anomalous dimensions arise from loop diagrams as in figure \ref{2pt}.
However, the bulk one-loop computation is notoriously difficult in higher spin gauge theory as is well known for gravity theory with spin two gauge field.

\begin{figure}
	\centering
	\includegraphics[keepaspectratio, scale=0.5]
	{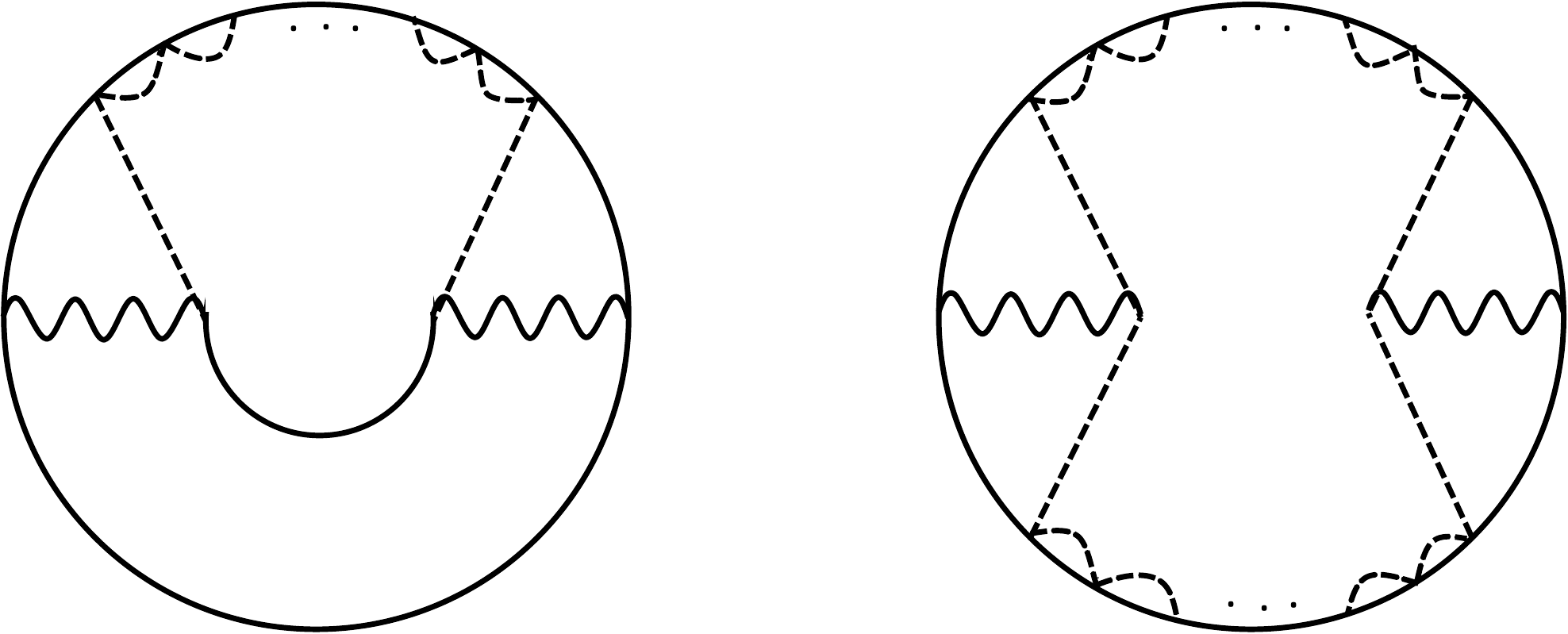}
	\caption{Bulk-to-bulk propagator along the loop may be replaced by the product of bulk-to-boundary operators and boundary two point functions.}
	\label{2ptmod}
\end{figure}

In \cite{Hikida:2016wqj,Hikida:2016cla} (based on previous works in \cite{Hartman:2006dy,Giombi:2011ya,Creutzig:2015hta}) the difficulty is evaded by utilizing the Witten's argument in \cite{Witten:2001ua} that deforming boundary theory by double-trace type operators is dual to changing the boundary condition of dual bulk fields. Once we admit the dictionary, the diagram with modified bulk-to-bulk propagators along the loop can be rewritten as a product of tree diagrams only with bulk-to-boundary propagators and boundary two point functions as in figure \ref{2ptmod}.
Furthermore, we can see that the bulk computation with tree Witten diagrams is equivalent to the boundary one in the conformal perturbation theory to the $1/N$ order. 
Therefore, what we have to do is to show that the modification of bulk-to-bulk propagators can be written only in terms of bulk-to-boundary operators and boundary two point functions without relying the AdS/CFT dictionary.
This was already done for the bulk scalar fields in \cite{Hartman:2006dy,Giombi:2011ya,Creutzig:2015hta}. 
For the holography with 3d O$(N)$ scalars, it was applied in \cite{Hikida:2016wqj}, and the same idea was already suggested in \cite{Giombi:2011ya}.
In this paper, we first repeat the analysis of \cite{Creutzig:2015hta} in a slightly different way as was done in \cite{Hartman:2006dy,Giombi:2011ya}, and then apply it to the case with bulk spinor fields.

The organization of this paper is as follows;
In the next section, we introduce the supersymmetric model with free bosons and fermions, and write down our notation for higher spin currents. Then we explain how to read off the anomalous dimensions form two point functions of higher spin currents in conformal perturbation theory.
In section \ref{sec:boson}, we compute the anomalous dimensions in the presence of deformation \eqref{def}. We first obtain them to the $\lambda^2$ order and then incorporate higher order corrections. In subsection \ref{aux}, we re-examine the model in the formulation with auxiliary fields. 
In section \ref{sec:fermion}, we study the case with another marginal deformation \eqref{defsusy}.
In section \ref{Bulk}, we derive the map from the one-loop diagram as in figure \ref{2pt} to a product of tree Witten diagrams as in figure \ref{2ptmod}. We first reproduce the known result with bulk scalar propagators and then extend the analysis to the case with bulk spinor propagators.
Section \ref{sec:conclusion} is devoted to conclusion and discussions.
In appendices, the detailed computations of Feynman integrals are given.

\section{Preliminary}
\label{Pre}

We examine supersymmetric U$(N)$ model in three dimensions consisting of 
$N$ complex scalars $\phi^i$ and Dirac fermions $\psi^i$ $(i=1,2,\cdots , N)$. 
The free action is given by
\begin{align}
S &= \int d^3 x \left[ \partial_\mu \bar{\phi}^i \partial^\mu \phi_i + \bar \psi^i \slashed{\partial} \psi_i \right] \, ,
\label{freeaction}
\end{align}
where the Gamma matrices satisfy $\{\gamma^\mu , \gamma^\nu \} = 2 g^{\mu \nu}$.
Two point functions are
\begin{align}
\langle \phi^{ i} (x_1) \bar{\phi}^{ j} (x_2) \rangle = C_\phi \frac{\delta^{ij}}{|x_{12}|} \, , \quad
\langle  \psi^{ i} (x_1) \bar \psi^j (x_2) \rangle = - C_\phi \slashed{\partial}_1 \frac{\delta^{ij}}{|x_{12}|} \, , \quad C_\phi = \frac{1}{4 \pi}
\end{align}
in the coordinate representation and 
\begin{align}
\langle \phi^i  (p) \bar{\phi}^j (-p) \rangle = \frac{\delta^{ij}}{|p|^2} \, , \quad
\langle \psi^i  (p) \bar{\psi}^j (-p) \rangle = - \frac{ i  \slashed{p}  \delta^{ij} }{|p|^2} 
\end{align}
in the momentum representation.
Here and in the following we adopt the formula of Fourier transform as
\begin{align}
 \Phi (p) = \int \frac{d^3 x}{(2 \pi)^{3/2}} \Phi (x) e^{- i p \cdot x} \, .
\end{align}
In the current case, we should set $\Phi =\phi^i$ or $\psi^i$.

The free boson and fermion theories have conserved currents $J_{\mu_1 \cdots \mu_s}$  and 
$\tilde J_{\mu_1 \cdots \mu_s}$ with traceless symmetric indices.
Introducing null polarization vector $\epsilon$ with $\epsilon \cdot \epsilon = 0$,
we denote $J^\epsilon_{s} (x) = J_{\mu_1 \cdots \mu_s} (x) \epsilon^{\mu_1} \cdots \epsilon^{\mu_s}$ and $\tilde J^\epsilon_{s} (x) = \tilde J_{\mu_1 \cdots \mu_s} (x) \epsilon^{\mu_1} \cdots \epsilon^{\mu_s}$. 
For the explicit forms, we use the conventions in \cite{Anselmi:1999bb} as
\begin{align}
J^\epsilon _s (x) = \sum_{k=0}^s a_k \hat \partial^k \bar{\phi}^{ i } \hat \partial^{s-k} \phi_i \, ,
\qquad a_k = \frac{(-1)^k}{2} \frac{\binom{s}{k} \binom{s-1}{k-1/2}}{\binom{s-1}{-1/2}} \, ,
\end{align}
and
\begin{align}
\tilde J^\epsilon _s (x) = \sum_{k=0}^{s-1} \tilde a_k \hat \partial^k \bar \psi^{ i } \hat  \gamma \hat \partial^{s-k-1} \psi_i \, ,
\qquad \tilde a_k = (-1)^k \frac{\binom{s-1}{k} \binom{s}{k+1/2}}{\binom{s}{1/2}} \, .
\end{align}
Here we have represented as
\begin{align}
\hat \partial \equiv \epsilon \cdot \partial \, , \quad
\hat \gamma \equiv \epsilon \cdot \gamma = \slashed{\epsilon} \, .
\end{align}
With the expressions, the two point functions are computed as (see \cite{Hikida:2016cla} and references therein)
\begin{align}
D_0^s \equiv\langle J^\epsilon_s (x_1)  J^\epsilon_s (x_2) \rangle 
= C_s \frac{(\hat x_{12})^{2s}}{(x_{12}^2)^{1+ 2 s}} \, , \quad
C_s = 2^{2s - 2 } N C_\phi^2 s \Gamma (2s)   \, ,
\end{align}
and 
\begin{align}
\tilde D_0^s \equiv\langle \tilde J^\epsilon_s (x_1)  \tilde J^\epsilon_s (x_2) \rangle 
= \tilde C_s \frac{(\hat x_{12})^{2s}}{(x_{12}^2)^{1 + 2 s}} \, , \quad
\tilde C_s =  2^{2s } N C_\phi^2 s^{-1} \Gamma (2s)    \, .
\end{align}
The relations
\begin{align}
\tilde C_s = 2^2 s^{-2} C_s \, \quad \text{or} \quad
\tilde D_0^s = 2^2 s^{-2} D_0^s 
\label{Csrel}
\end{align}
will be used for later analysis.

We deform the free action \eqref{freeaction} by adding
\begin{align}
\Delta_{\lambda ,\kappa} S = \Delta_\lambda S + \Delta_\kappa S \, ,
\end{align}
where $\Delta_\lambda S$ and $\Delta_\kappa S$ are given in \eqref{def} and \eqref{defsusy}, respectively. We treat the additional part in conformal perturbation theory as
\begin{align}
\left \langle \prod_{a=1}^n \Phi_a (x_a) \right \rangle_{\lambda ,\kappa}
 = \frac{\left \langle \prod_{a=1}^n \Phi_a (x_a) e^{- \Delta_{\lambda , \kappa} S}\right \rangle_{0}}{\left \langle e^{- \Delta_{\lambda , \kappa} S}\right \rangle_{0}} \, .
 \label{modcorr}
\end{align}
Here $\Phi_a$ is an operator and the right hand side is computed in the free theory.

In the presence of deformations, the higher spin currents $J^\epsilon_s (x)$ and 
$\tilde J^\epsilon_s (x)$ are not conserved anymore and possess anomalous dimensions.
We read off them from the two point functions of higher spin currents by using the modified correlators as in \eqref{modcorr}.
In the current case, there are two types of higher spin currents, and we would meet a diagonalization problem to obtain the anomalous dimensions, see, e.g.,  \cite{Gaberdiel:2015uca}.
Denoting two independent linear combinations as $J^{\epsilon, \alpha}_s (x)$ with $\alpha = 1,2$, the two point functions are of the forms
\begin{align}
\langle J^{\epsilon , \alpha}_s (x_1)  J^{\epsilon , \beta}_s (x_2) \rangle_{\lambda, \kappa}
&= N_s^{\alpha \beta} \frac{(\hat x_{12})^{2s}}{(x_{12}^2)^{1 + 2s + \gamma_s^{\alpha \beta}}} \, , \label{2ptbefore}
\end{align}
which are fixed by the symmetry.
Here we have used $\hat x_{12} = \epsilon \cdot x_{12}$.
We may change the basis for $J^{\epsilon , \alpha}_s (x)$ such that 
\begin{align}
N_s^{\alpha \beta} = \delta^{\alpha \beta} + n_s^{\alpha \beta} + \mathcal{O}(N^{-2}) \, , \quad 
N_s^{\alpha \beta} \gamma_s^{\alpha \beta} = \delta^{\alpha \beta} \gamma_s^\alpha + \mathcal{O} (N^{-2} )
\label{diag}
\end{align}
with $n_s^{\alpha \beta}, \gamma_s^\alpha$ of order $1/N$.
Then the two point functions become
\begin{align}
\nonumber
&\langle J^{\epsilon , \alpha}_s (x_1)  J^{\epsilon , \beta}_s (x_2) \rangle_{\lambda, \kappa} \\
&=  \delta^{\alpha \beta} \frac{(\hat x_{12})^{2s}}{(x_{12}^2)^{1 + 2s }}
- \delta^{\alpha \beta}  \gamma_s^{\alpha } \frac{(\hat x_{12})^{2s}}{(x_{12}^2)^{1 + 2s }} \log (x_{12}^2) 
+ n_s^{\alpha \beta} \frac{(\hat x_{12})^{2s}}{(x_{12}^2)^{1 + 2s }}
+ \mathcal{O}(N^{-2}) \, . 
\end{align}
Therefore, the anomalous dimension $\gamma^\alpha_s$ of  $J^{\epsilon, \alpha}_s$ to the $1/N$ order can be read off from the factor in front of $\log (x_{12}^2)$.

\section{Deformation with bosonic operators}
\label{sec:boson}

In this section we deform the supersymmetric model by \eqref{def} and compute the anomalous dimensions of higher spin currents. We first study them to the first non-trivial order in the deformation parameter $\lambda$ in the next subsection and then systematically include higher order corrections in subsection \ref{higher}.
In subsection \ref{aux}, we rewrite the deformed action by introducing auxiliary fields, and examine the relations to previous works on the critical O$(N)$ scalars and the Gross-Neveu model such as in \cite{Diab:2016spb,Hikida:2016cla}.

\subsection{Anomalous dimensions to the first non-trivial order}

We deform the free theory by the marginal deformation in \eqref{def}. 
The two point functions of the scalar operators $\mathcal{O}(x) = \bar{\phi}^{i} \phi_i$ and $\tilde{\mathcal{O}}(x) = \bar \psi^i \psi_i$ at $\lambda = 0$ are given by
\begin{align}
\langle \mathcal{O}(x_1) \mathcal{O} (x_2) \rangle _0 =  \frac{N C_\phi^2}{x_{12}^2} \, ,
\quad
\langle \tilde{\mathcal{O}}(x_1) \tilde{\mathcal{O}} (x_2) \rangle _0 =  \frac{2 N C_\phi^2 }{(x_{12}^2)^2} \, .
\label{2ptscalars}
\end{align}
We would like to compute the two point functions of higher spin currents in the presence of marginal deformation. The modified correlators in \eqref{modcorr} to the first non-trivial orders both in $\lambda$ and $1/N$ are
\begin{align}
& \langle J^\epsilon_s (x_1) J^\epsilon_s (x_2) \rangle_\lambda
 = \langle J^\epsilon_s (x_1) J^\epsilon_s (x_2) \rangle_0  + \lambda^2 I_1  + \mathcal{O} (\lambda^4) \, , \label{JJ}\\
 &\langle \tilde J^\epsilon_s (x_1) \tilde J^\epsilon_s (x_2) \rangle_\lambda
   = \langle \tilde J^\epsilon_s (x_1) \tilde J^\epsilon_s (x_2) \rangle_0  + \lambda^2 I_2 + \mathcal{O} (\lambda^4) \, , \label{tJtJ}\\ 
 &\langle J^\epsilon_s (x_1) \tilde J^\epsilon_s (x_2) \rangle_\lambda
   = \lambda^2 I_3 + \mathcal{O} (\lambda^4) \, . \label{JtJ}
 \end{align}
Here $I_1$, $I_2$, and $I_3$  are defined by the following integrals as
\begin{align}
& I_1 = \frac12 \int d^3 x_3 d^3 x_4 
  \langle  J^\epsilon_s (x_1) J^\epsilon_s (x_2) \mathcal{O} (x_3) \mathcal{O} (x_4) \rangle_0
  \langle \tilde{\mathcal{O}} (x_3) \tilde{\mathcal{O}} (x_4) \rangle_0 \, , \label{I1} \\
 & I_2 = \frac12 \int d^3 x_3 d^3 x_4 
    \langle \tilde J^\epsilon_s (x_1) \tilde J^\epsilon_s (x_2) \tilde{\mathcal{O}} (x_3) \tilde{\mathcal{O}} (x_4) \rangle_0     \langle \mathcal{O} (x_3) \mathcal{O}(x_4) \rangle_0\, , \label{I2} \\ 
 & I_3 =\frac12 \int d^3 x_3 d^3 x_4 
   \langle J^\epsilon_s (x_1)  \mathcal{O} (x_3) \mathcal{O}(x_4) \rangle_0 
       \langle \tilde J^\epsilon_s (x_2)  \tilde{\mathcal{O}} (x_3) \tilde{\mathcal{O}} (x_4) \rangle_0
 \, . \label{I3}
\end{align}
In order to read off the anomalous dimensions, we need to extract terms proportional to $\log(x_{12}^2)$ from the integrals.
Such terms from the above integrals have been already obtained in \cite{Hikida:2016wqj,Hikida:2016cla}, so we can just borrow the results there.
See appendix \ref{app:bosonic} for details.

As explained in the previous section, we need to solve the diagonalization problem to obtain the anomalous dimensions by choosing a proper linear combination of currents to satisfying the conditions \eqref{diag}.  In the present case, the proper choice turns out to be
\begin{align}
J^{\epsilon ,\pm}_s = \frac{1}{\sqrt{2 C_s}} J^{\epsilon}_s \pm \frac{1}{\sqrt{2 \tilde C_s}}\tilde J^\epsilon_s \, .
\label{linear}
\end{align}
The two point functions of these currents are
\begin{align}
&\langle J^{\epsilon ,\pm}_s (x_1) J^{\epsilon ,\pm}_s (x_2) \rangle_\lambda \nonumber \\
& \qquad = \langle J^{\epsilon ,\pm}_s (x_1) J^{\epsilon ,\pm}_s (x_2) \rangle_0 
+ \lambda^2 \left[ \frac{1}{2C_s} I_1 + \frac{1}{2 \tilde C_s}  I_2 \pm \frac{s}{2 C_s} I_3 \right]
+ \mathcal{O}(\lambda^4) \, ,
\end{align}
where we have used a relation in \eqref{Csrel}.
Using the results in appendix \ref{app:bosonic},
the anomalous dimensions to the $\lambda^2$ order and to the leading order in the $1/N$-expansion are read off as%
\footnote{As in \eqref{tlambda} below, a convenient parameter is not $\lambda$ but $\tilde \lambda = N \lambda / 8$. Therefore, we take large $N$ but keep $\tilde \lambda$ finite. In this sense, the expressions in \eqref{annboson} are of the $1/N$ order and of the $\tilde \lambda^2$ order.}
\begin{align}
 \gamma_s^\pm  & =  \frac{\lambda^2  N}{12 \pi^2} \frac{(s-1)(s+1)}{(2s-1)(2s+1)}   \mp \mathcal{P}_s \frac{\lambda^2  N}{8 \pi^2} \frac{s}{(2s - 1 ) (2s+1)} 
 + \mathcal{O} (\lambda^4)
 \label{annboson}
\end{align}
with 
\begin{align}
\mathcal{P}_s = \frac{1 + (-1)^s }{2} \, .
\end{align}
Thus we have 
\begin{align}
\gamma_s^\pm  =  \frac{\lambda^2  N}{12 \pi^2} \frac{(s-1)(s+1)}{(2s-1)(2s+1)}  + \mathcal{O} (\lambda^4)
\label{anoodd}
\end{align}
for odd $s$ and
\begin{align}
\gamma_s^\pm  = \frac{\lambda^2 N }{24 \pi^2} \frac{s \mp 2}{2s \mp 1}   + \mathcal{O} (\lambda^4)
\label{anoeven}
\end{align}
for even $s$. Here $J^{\epsilon,\pm}_1$ correspond to spin 1 currents for global symmetry $U(1) \otimes U(1) \in U(N) \otimes U(N)$, which are not broken by the deformation as $\gamma^\pm_1 = 0$.
Moreover, $J^{\epsilon,+}_2$ corresponds to the energy momentum tensor, and hence we have $\gamma_2^+ = 0$. 

\subsection{Higher order corrections}
\label{higher}

We would like to move to the higher order corrections in the $\lambda$-expansion.
A large part of corrections can be incorporated simply by replacing the parameter $\lambda^2$ in \eqref{anoodd} and \eqref{anoeven} by, say, $\lambda^2 / (1 + \tilde \lambda^2)$ with
\begin{align}
\label{tlambda}
\tilde \lambda = \frac{N}{8} \lambda \, .
\end{align}
However, not all of the corrections can be treated in this way, and new integrals should be evaluated as well.

A type of correction can be summarized by the replacements of 
$\langle \tilde{\mathcal{O}}(x_3) \tilde{\mathcal{O}}(x_4) \rangle_0$ in \eqref{I1} by
$\langle \tilde{\mathcal{O}}(x_3) \tilde{\mathcal{O}}(x_4) \rangle_\lambda$ and
$\langle \mathcal{O}(x_3) \mathcal{O}(x_4) \rangle_0$ in \eqref{I2} by
$\langle \mathcal{O}(x_3) \mathcal{O}(x_4) \rangle_\lambda$.
In the momentum representation, the two point functions \eqref{2ptscalars} at $\lambda = 0$ become
\begin{align}
G(p) = \langle \mathcal{O}(p) \mathcal{O} (-p) \rangle _0 =  \frac{N}{8} \frac{1}{ |p|} \, ,
\quad
\tilde G(p) = \langle \tilde{\mathcal{O}}(p) \tilde{\mathcal{O}} (-p) \rangle _0  = - \frac{N}{8} | p | \, .
\end{align}
Therefore, at finite $\lambda$, the two point functions become 
\begin{align}
\nonumber
\langle \mathcal{O}(p) \mathcal{O} (-p) \rangle _\lambda &= 
G(p)  + \lambda^2 G(p) \tilde G(p) G(p) + \lambda^4 G(p) \tilde G(p) G(p) \tilde G(p) G(p) + \cdots 
\\
&= G(p) (1 - \tilde{\lambda}^2 + \tilde{\lambda}^4 - \cdots ) = G(p) \frac{1}{ 1 + \tilde{\lambda}^2 }
\, ,
\end{align}
and similarly
\begin{align}
\nonumber
\langle \tilde{\mathcal{O}}(p) \tilde{\mathcal{O}} (p) \rangle _\lambda = 
\tilde G(p) \frac{1}{ 1 + \tilde{\lambda}^2 }
\, .
\end{align}
Therefore, the corrections of this type simply replace $\tilde \lambda^2$ by 
$\tilde{\lambda}^2 /(1 + \tilde{\lambda}^2)$.
For \eqref{I3} we may rewrite as
\begin{align}
\nonumber
 \lambda^2  I_3 =\frac{\lambda^2}{2} \int d^3 x_3 d^3 x_4 d^3 x_5 d^3 x_6 &
   \langle J^\epsilon_s (x_1)  \mathcal{O} (x_3) \mathcal{O}(x_4) \rangle_0 
       \langle \tilde J^\epsilon_s (x_2)  \tilde{\mathcal{O}} (x_5) \tilde{\mathcal{O}} (x_6) \rangle_0 \\
      & \qquad \times
       \delta^{(3) } (x_{35}) \delta^{(3) } (x_{46}) 
 \, . 
\end{align}
Higher order corrections modify the delta functions (or the identities in the momentum representation) as
\begin{align}
1 + \lambda^2 G(p) \tilde G(p) + \lambda^4 G(p) \tilde G(p) G(p) \tilde G(p) + \cdots
= \frac{1}{1 + \tilde \lambda^2} \, .
\end{align}
This means that we should replace  each $\tilde{\lambda}$ by 
  $ \tilde{\lambda} /(1 + \tilde{\lambda}^2)$, i.e.,  $\tilde{\lambda}^2$ by
  $ (\tilde{\lambda} /(1 + \tilde{\lambda}^2))^2$.

At the $\lambda^4$ order, we can see that the following new types of integral appear as
\begin{align}
\label{I4}
  I_4 =\frac{1}{2} \int d^3 x_3 d^3 x_4 d^3 x_5 d^3 x_6 &
   \langle J^\epsilon_s (x_1)  \mathcal{O} (x_3) \mathcal{O}(x_4) \rangle_0 
       \langle J^\epsilon_s (x_2)  \mathcal{O} (x_5) \mathcal{O} (x_6) \rangle_0
  \\ & \qquad  \times
   \langle \tilde{\mathcal{O}} (x_3)  \tilde{\mathcal{O}} (x_5) \rangle_0 
    \langle \tilde{\mathcal{O}} (x_4)  \tilde{\mathcal{O}} (x_6) \rangle_0 
 \, ,  \nonumber \\
 \label{I5}
I_5 =\frac{1}{2} \int d^3 x_3 d^3 x_4 d^3 x_5 d^3 x_6 &
   \langle \tilde J^\epsilon_s (x_1)  \tilde{\mathcal{O}} (x_3) \tilde{\mathcal{O}}(x_4) \rangle_0 
       \langle \tilde J^\epsilon_s (x_2)  \tilde{\mathcal{O}} (x_5) \tilde{\mathcal{O}} (x_6) \rangle_0
  \\ & \qquad  \times
   \langle \mathcal{O} (x_3)  \mathcal{O} (x_5) \rangle_0 
    \langle \mathcal{O} (x_4)  \mathcal{O} (x_6) \rangle_0 \, .  \nonumber
 \end{align}
 The integrals have been already evaluated in \cite{Hikida:2016cla}, and we use the results, see appendix \ref{app:bosonic}.
Further higher order corrections in $\lambda$ replace
 $\langle \tilde{\mathcal{O}}(x_3) \tilde{\mathcal{O}}(x_4) \rangle_0$ with
  $\langle \tilde{\mathcal{O}}(x_3) \tilde{\mathcal{O}}(x_4) \rangle_\lambda$ in \eqref{I4}
  (and similarly for \eqref{I5}) as before.

Summarizing the above analysis, the two point functions of higher spin currents at finite $\lambda$ can be given by 
\begin{align}
&\langle J^{\epsilon ,\pm}_s (x_1) J^{\epsilon ,\pm}_s (x_2) \rangle_\lambda
 = \langle J^{\epsilon ,\pm}_s (x_1) J^{\epsilon ,\pm}_s (x_2) \rangle_0 
 + \left( \frac{8}{N} \right)^2 \frac{\tilde{\lambda}^2}{1 + \tilde{\lambda}^2 } 
  \left[ \frac{1}{2C_s} I_1 + \frac{1}{2 \tilde C_s}  I_2 \right] \nonumber  \\
 & \pm \left( \frac{8}{N} \right)^2 \left( \frac{\tilde{\lambda}}{1 + \tilde{\lambda}^2 } \right)^2  \frac{s}{2 C_s}  I_3  + \left( \frac{8}{N} \right)^4 \left( \frac{\tilde{\lambda}^2}{1 + \tilde{\lambda}^2 } \right)^2    \left[ \frac{1}{2C_s} I_4 + \frac{1}{2 \tilde C_s} I_5 \right]
 \, , \label{Fulllambda}
\end{align}
where we can still use the linear combinations in \eqref{linear}.
Referring to the results in appendix \ref{app:bosonic},  the anomalous dimensions to the $1/N$ order are thus%
\footnote{In CFTs with higher spin symmetry broken in $1/N$, broken symmetry still provides quite strong constraints on correlators as shown in \cite{Maldacena:2012sf}. The dependence on the deformation parameter $\tilde \lambda$ in \eqref{anomalousb} may be explained by using broken symmetry as was done in \cite{Giombi:2016zwa} for similar parameter dependence in Chern-Simons matter theories. The same argument can be applied to \eqref{anomalousf} as well.}
\begin{align}
\nonumber
 \gamma_s^\pm  & =  \frac{16}{3 N \pi^2}  \frac{\tilde{\lambda}^2}{1 + \tilde{\lambda}^2 } 
 \frac{(s-1)(s+1)}{(2s-1)(2s+1)} \\
   & \qquad + \mathcal{P}_s \frac{8 }{N \pi^2} 
 \left[ \mp \left( \frac{\tilde{\lambda}}{1 + \tilde{\lambda}^2 }  \right) ^2
  - \left( \frac{\tilde{\lambda}^2}{1 + \tilde{\lambda}^2 } \right)^2  \right]\frac{s}{(2s-1)(2s+1)}  \, .
  \label{anomalousb}
\end{align}
This is a main finding in this paper.
Setting $s=1$, we obtain $\gamma_1^\pm =0$, thus the global 
$\text{U}(1) \otimes \text{U}(1) \in \text{U}(N) \otimes \text{U}(N)$ symmetry is not broken to the full orders in $\tilde \lambda$.
For $\gamma_s^+$ with even $s$, we can rewrite as
\begin{align}
\gamma_s^+ = \frac{8}{3 N \pi^2}  \frac{\tilde{\lambda}^2}{1 + \tilde{\lambda}^2 } \frac{s-2}{2s-1} \, .
\end{align}
In particular, we have $\gamma_2^+ = 0$, and this means that the the deformation \eqref{def} is exactly marginal to the $1/N$ order.

It is useful to examine the results \eqref{anomalousb} in the limit $\tilde \lambda \to \infty$,
which are given as
\begin{align}
\label{anomalousblimit}
\gamma_s^\pm \to  \frac{16}{3 N \pi^2}  \frac{(s-1)(s+1)}{(2s-1)(2s+1)} 
- \mathcal{P}_s \frac{8}{N \pi^2} \frac{s}{(2s-1)(2s+1)} \, .
\end{align}
The limit of supersymmetric model should correspond to the sum of the critical U$(N)$ vector model and the Gross-Neveu model. This can be seen both from CFT viewpoint as in the next subsection and from  the dual gravity theory as in \cite{Leigh:2003gk,Sezgin:2003pt}, see also section \ref{Bulk}. 
The anomalous dimensions for the critical U$(N)$ model can be found in \cite{Hikida:2016cla} with a slight modification of O$(N)$ case  in \cite{Lang:1992zw}, and those for  the Gross-Neveu model 
were computed in \cite{Muta:1976js}. See \cite{Skvortsov:2015pea,Giombi:2016hkj,Manashov:2016uam} for recent related works.
Comparing these results and \eqref{anomalousblimit}, we find agreements.

\subsection{Introducing auxiliary fields}
\label{aux}

In this subsection, we describe the deformation of supersymmetric model by \eqref{def} 
as 
\begin{align}
S
= \int d^3 x \left[ \partial_\mu \bar{\phi}^i \partial^\mu \phi_i + \bar \psi^i \slashed{\partial} \psi_i  + \tilde{\sigma} \mathcal{O} + \sigma \tilde{\mathcal{O}}  - \frac{1}{ \lambda } \sigma \tilde \sigma \right]  \, ,
\label{actionv2}
\end{align}
where we have introduced auxiliary fields $\sigma, \tilde{\sigma} $.
We may study the model by treating the last term in \eqref{actionv2} perturbatively. 
It is useful to move to this formulation for the purpose to see more direct relations to previous works as in \cite{Diab:2016spb,Hikida:2016cla}.
This is because the last term in \eqref{actionv2} vanishes at $\lambda \to \infty$ limit, where the theory reduces to the sum of the critical U$(N)$ model and the Gross-Neveu model.

Using the large $N$ factorization of $\mathcal{O}$ and $\tilde{\mathcal{O}}$, the effective propagators for $\sigma , \tilde \sigma$ become (see, e.g., \cite{Diab:2016spb})
\begin{align}
&G_\sigma (p) = \langle \sigma (p) \sigma (-p) \rangle_0
=  (- \langle \mathcal{O} (p) \mathcal{O} (-p) \rangle_0 )^{-1} = - \frac{8}{N} |p| \, , \\
&G_{\tilde{\sigma}}(p) = \langle \tilde \sigma (p) \tilde \sigma (-p) \rangle _0 
=  (- \langle \tilde{\mathcal{O}} (p) \tilde{\mathcal{O}} (-p) \rangle_0 )^{-1} = \frac{8}{N} \frac{1}{|p|} \, .
\end{align}
At the  limit $\lambda \to \infty$, corrections to the two point functions of higher spin currents come from the integrals $I_1$, $I_2$, $I_4$, $I_5$ in \eqref{I1}, \eqref{I2}, \eqref{I4}, \eqref{I5} but with $\langle \tilde{\mathcal{O}} \tilde{\mathcal{O}} \rangle_0, \langle  \mathcal{O} \mathcal{O} \rangle_0$ replaced by $\langle \sigma \sigma \rangle_0 , \langle \tilde \sigma \tilde  \sigma \rangle_0$. This can be seen from \eqref{Fulllambda} at $\tilde \lambda \to \infty$, for instance. The regularization with $\Delta$ used in the appendix \ref{app:bosonic} may be understood by the shifts of exponents as
\begin{align}
&G_\sigma (p) = \langle \sigma (p) \sigma (-p) \rangle _0 = - \frac{8}{N} \frac{1}{(p^2)^{-1/2 + \Delta_1}} \, , \\
&G_{\tilde{\sigma}}(p) = \langle \tilde \sigma (p) \tilde \sigma (-p) \rangle _0  =  \frac{8}{N} \frac{1}{(p^2)^{1/2 +\Delta_2}} \, ,
\end{align}
where we should set $\Delta_1$, $\Delta_2$ properly.

Higher order corrections in $1/\lambda$ can be understood also in this formulation. Here we set $\Delta_1 + \Delta_2 = 0$ for simplicity, but it is not so difficult to work with $\Delta_1 + \Delta_2 \neq 0$.
Due to the last term in \eqref{actionv2}, the effective propagators should be corrected as
\begin{align}
& \langle \sigma (p) \sigma (-p) \rangle _{1/\lambda}
  = G_\sigma (p) + \lambda^{-2} G_\sigma (p) G_{\tilde \sigma} (p)   G_\sigma (p) +\cdots
   = \frac{1}{1 + \tilde \lambda^{-2}} G_\sigma (p) \, , \\
& \langle \tilde{\sigma} (p) \tilde{\sigma} (-p) \rangle _{1/\lambda}
 = G_{\tilde{\sigma}} (p) + \lambda^{-2} G_{\tilde{\sigma}} (p) G_{\sigma} (p)   G_{\tilde{\sigma}} (p) +\cdots
 = \frac{1}{1 + \tilde \lambda^{-2}} G_{\tilde{\sigma}} (p) \, .
\end{align}
Replacing $\langle \sigma \sigma \rangle_0 , \langle \tilde \sigma \tilde  \sigma \rangle_0$
by $\langle \sigma \sigma \rangle_{1/\lambda} , \langle \tilde \sigma \tilde  \sigma \rangle_{1/\lambda}$, we reproduce the previous result in \eqref{Fulllambda} except for the contribution arising from the integral of $I_3$-type.

The integral of $I_3$-type contributes from the $1/\lambda^{2}$ order. 
This type of correction is absent for the model made from simply the sum of the critical U$(N)$ scalars and the Gross-Neveu model.
At the order, the integral can be written as%
\footnote{The shifts by $\Delta$ for the integral $I_3$ may be interpreted as the extra shifts introduced in section 3.3 of \cite{Hikida:2016cla}.}
\begin{align}
  I_3 ' = \frac{1}{2 \lambda^2}\int \prod_{n=3}^8 d^3 x_n & 
 \langle J^\epsilon_s (x_1)  \mathcal{O} (x_3) \mathcal{O}(x_4) \rangle_0 
 \langle \tilde J^\epsilon_s (x_2)  \tilde{\mathcal{O}} (x_7) \tilde{\mathcal{O}} (x_8) \rangle_0   \\
  & \times    \langle \sigma (x_3) \sigma (x_5) \rangle_0  \langle \tilde{\sigma} (x_5) \tilde{\sigma} (x_7)   \rangle_0 
  \langle \sigma (x_4) \sigma (x_6) \rangle_0  
  \langle \tilde{\sigma} (x_6) \tilde{\sigma} (x_8)   \rangle_0 
 \, .  \nonumber 
\end{align}
For $\Delta_1 + \Delta_2 = 0$, we have
\begin{align}
\int d^3 x_5 \langle \sigma (x_3) \sigma (x_5) \rangle_0  \langle \tilde{\sigma} (x_5) \tilde{\sigma} (x_7)   \rangle_0 = - \left( \frac{8}{N} \right)^2 \delta^{(3)} (x_{37}) \, ,
\end{align}
and similarly for the $x_6$-integral. Using these delta functions, we can see that $I_3 '$ reduces to $I_3$. The higher order corrections can be included by replacing $1$ by $1/(1 + \tilde \lambda^{-2})$ as shown above. Noticing
\begin{align}
\frac{1}{\lambda^2} \left( \frac{1}{1 + \tilde \lambda^{ - 2} } \right)^2
 = \left( \frac{N}{8} \right)^2  \left( \frac{\tilde \lambda}{1 + \tilde \lambda^{ 2} } \right)^2 \, ,
\end{align}
the contribution from this type of integral in \eqref{Fulllambda} can be also reproduced.

\section{Deformation with fermionic operators}
\label{sec:fermion}

The supersymmetric model admits another marginal deformation as in \eqref{defsusy} along with the one in \eqref{def}, which was examined in the previous section. 
In this section, we compute the anomalous dimensions of higher spin currents in the deformed theory by \eqref{defsusy}. We may use both the deformations simultaneously such as the supersymmetric one with $2 \lambda = \kappa$. In that case we just need to add each contribution to anomalous dimensions.

The deformation \eqref{defsusy} consists of spinor operator $\mathcal{K}  = \bar{\phi}^{i} \psi_i$ $(\bar{\mathcal{K}}= \bar \psi^i \phi_i)$, whose two point function is
\begin{align}
\langle \mathcal{K} (x_1) \mathcal{\bar K} (x_2) \rangle _0 
=  N C_\phi^2 \frac{\slashed{x}_{12}}{(x_{12}^2)^2}
 = - \frac{N C_\phi^2}{2}\slashed{\partial}_1 \frac{1}{x_{12}^2 } 
\label{2ptsfermions}
\end{align}
at $\kappa = 0$.
We compute the two point functions of higher spin currents in the presence of marginal deformation.
The modified correlators \eqref{modcorr} to the first non-trivial orders both in $\kappa$ and $1/N$  are
\begin{align}
& \langle J^\epsilon_s (x_1) J^\epsilon_s (x_2) \rangle_\kappa
 = \langle J^\epsilon_s (x_1) J^\epsilon_s (x_2) \rangle_0  + \kappa^2 (\tilde I_1 + \tilde I_2)  + \mathcal{O} (\kappa^4) \, , \label{JJf}\\
 &\langle \tilde J^\epsilon_s (x_1) \tilde J^\epsilon_s (x_2) \rangle_\kappa
   = \langle \tilde J^\epsilon_s (x_1) \tilde J^\epsilon_s (x_2) \rangle_0 + \kappa^2 (\tilde I_3+ \tilde I_4)  + \mathcal{O} (\kappa^4) \, , \label{tJtJf}\\ 
 &\langle J^\epsilon_s (x_1) \tilde J^\epsilon_s (x_2) \rangle_\kappa
   = \kappa^2 (\tilde I_5+ \tilde I_6)  + \mathcal{O} (\kappa^4) \, . \label{JtJf}
 \end{align}
 Here $\tilde I_a$ $(a=1,\ldots,6)$ are integrals defined as
\begin{align}
& \tilde I_1 = -  \int d^3 x_3 d^3 x_4 
  \langle  J^\epsilon_s (x_1) J^\epsilon_s (x_2) \mathcal{K}(x_3) \bar{\mathcal{K}} (x_4) \rangle_0
  \langle \mathcal{K} (x_4) \mathcal{\bar K} (x_3) \rangle_0 \, , \label{tI1} \\
 & \tilde I_2 = - \int d^3 x_3 d^3 x_4 
   \langle J^\epsilon_s (x_1)  \mathcal{K} (x_3) \mathcal{\bar K}(x_4)  \rangle_0 
       \langle J^\epsilon_s (x_2) {\mathcal{K}} (x_4) {\mathcal{\bar K}} (x_3)  \rangle_0  
 \, , \label{tI2} \\
 & \tilde I_3 = -  \int d^3 x_3 d^3 x_4 
   \langle  \tilde J^\epsilon_s (x_1) \tilde J^\epsilon_s (x_2) \mathcal{K} (x_3) \bar{\mathcal{K}} (x_4) \rangle_0
   \langle \mathcal{K} (x_4) \mathcal{\bar K} (x_3) \rangle_0 \, , \label{tI3} \\
  & \tilde I_4 = - \int d^3 x_3 d^3 x_4 
    \langle \tilde J^\epsilon_s (x_1)  \mathcal{K}(x_3) \mathcal{\bar K} (x_4)  \rangle_0 
        \langle \tilde J^\epsilon_s (x_2) {\mathcal{K}} (x_4) {\mathcal{\bar K}} (x_3)  \rangle_0  
  \, , \label{tI4} \\
  & \tilde I_5 = -  \int d^3 x_3 d^3 x_4 
    \langle  J^\epsilon_s (x_1) \tilde J^\epsilon_s (x_2) \mathcal{K} (x_3) \bar{\mathcal{K}}  (x_4) \rangle_0
    \langle \mathcal{K} (x_4) \mathcal{\bar K}  (x_3) \rangle_0 \, , \label{tI5} \\
   & \tilde I_6= - \int d^3 x_3 d^3 x_4 
     \langle J^\epsilon_s (x_1)  \mathcal{K} (x_3) \mathcal{\bar K}  (x_4)  \rangle_0 
         \langle \tilde J^\epsilon_s (x_2) {\mathcal{K}} (x_4) {\mathcal{\bar K}} (x_3)  \rangle_0  
   \, . \label{tI6}
\end{align}
These integrals actually coincide with those evaluated in \cite{Hikida:2016cla}, so we can utilize the results obtained there as summarized in appendix \ref{app:fermionic}.
The anomalous dimensions of higher spin currents can be computed as in the previous section, and the results happen to be the same as \eqref{annboson} but $\lambda$ replaced by $\kappa$.

Higher order corrections in $\kappa$ can be included as in subsection \ref{higher}.
One type of corrections can be incorporated by replacing $ \langle \mathcal{K} (x_4) \mathcal{\bar K}  (x_3) \rangle_0$ in $\tilde I_1$, $\tilde I_3$, $\tilde I_5$ by $ \langle \mathcal{K} (x_4) \mathcal{\bar K}  (x_3) \rangle_\kappa$.
In the momentum representation, the two point function \eqref{2ptsfermions} becomes
\begin{align}
F(p) \equiv \langle \mathcal{K} (p) \bar{\mathcal{K}} (-p) \rangle_0 = - \frac{N}{16} \frac{i \slashed{p}}{|p|} \, .
%\quad \mathcal{K} (p) \equiv \int \frac{d^3 x} {(2\pi)^{3/2}} \mathcal{K} (x) e^{- i p \cdot x} 
\end{align}
Thus the two point function receives corrections at finite $\kappa$ as
\begin{align}
 \langle \mathcal{K} (p) \bar{\mathcal{K}} (-p) \rangle_\kappa =
F(p) + \kappa F(p)^2 + \kappa^2 F(p)^3 + \cdots 
 = F(p,\kappa)^o +  F(p,\kappa)^e \, .
\end{align}
Here $F(p,\kappa)^o $ linearly depends on the gamma matrix but $F(p,\kappa)^e$ does not as
\begin{align} 
 F(p,\kappa)^o =
 F(p) \frac{1}{1 + \tilde{\kappa}  ^2} \, , \quad F(p,\kappa)^e =
 \frac{N}{16}\frac{\tilde{\kappa} }{1 + \tilde{\kappa} ^2} \, , \quad  \tilde{\kappa} = \frac{N}{16} \kappa \, .
 \label{tkappa}
\end{align}
Noticing that trace over the odd number of gamma matrix vanishes,
corrections to the integrals $\tilde I_1$, $\tilde I_3$, $\tilde I_5$ can be included by replacing $F(p)$ by $ F(p,\kappa)^o $. 
Therefore we should replace $\tilde{\kappa} ^2 $ by $\tilde{\kappa} ^2 / (1 + \tilde{\kappa} ^2)$.
Similarly,  for $\tilde I_2$, $\tilde I_4$, $\tilde I_6$, we replace $1$ in the momentum representation by
\begin{align}
1 + \kappa^2 F(p)^2 + \kappa^4 F(p)^4 + \cdots = \frac{1}{1 + \tilde{\kappa} ^2} \, .
\end{align} 
This means that we should replace $\tilde{\kappa} ^2 $ by $ (\tilde{\kappa}  / (1 + \tilde{\kappa}  ^2))^2$.

At the order of ${\kappa }^4$, there are new types of correction appearing as
\begin{align}
\label{tI7} 
  \tilde I_{7} = - \int d^3 x_3 d^3 x_4 d^3 x_5 d^3 x_6 &
   \langle J^\epsilon_s (x_1)  \mathcal{K}(x_3) \mathcal{\bar K}(x_4)  \rangle_0 
           \langle \mathcal{K} (x_4) \mathcal{\bar K} (x_5) \rangle_0 
       \\ \nonumber & \qquad \times
       \langle J^\epsilon_s (x_2) {\mathcal{K}} (x_5) {\mathcal{\bar K}}(x_6)  \rangle_0  
            \langle \mathcal{K} (x_6) \mathcal{\bar K} (x_3) \rangle_0 
 \, , \\
\label{tI8} 
 \tilde  I_{8} = - \int d^3 x_3 d^3 x_4 d^3 x_5 d^3 x_6 &
   \langle \tilde J^\epsilon_s (x_1)  \mathcal{K}(x_3) \mathcal{\bar K}(x_4)  \rangle_0 
           \langle \mathcal{K} (x_4) \mathcal{\bar K} (x_5) \rangle_0 
       \\ \nonumber & \qquad \times
       \langle \tilde J^\epsilon_s (x_2) {\mathcal{K}} (x_5) {\mathcal{\bar K}}(x_6)  \rangle_0  
            \langle \mathcal{K} (x_6) \mathcal{\bar K} (x_3) \rangle_0 
 \, , \\
 \label{tI9} 
  \tilde  I_{9} = - \int d^3 x_3 d^3 x_4 d^3 x_5 d^3 x_6 &
    \langle J^\epsilon_s (x_1)  \mathcal{K}(x_3) \mathcal{\bar K}(x_4)  \rangle_0 
            \langle \mathcal{K} (x_4) \mathcal{\bar K} (x_5) \rangle_0 
        \\ \nonumber & \qquad \times
        \langle \tilde J^\epsilon_s (x_2) {\mathcal{K}} (x_5) {\mathcal{\bar K}}(x_6)  \rangle_0  
             \langle \mathcal{K} (x_6) \mathcal{\bar K} (x_3) \rangle_0 
  \, .
\end{align}
Further higher order corrections in $\kappa$ can be included by replacing $\tilde{\kappa} ^4 $ with
$ (\tilde{\kappa}^2/(1 + \tilde{\kappa}^2))^2$.
These integrals have not appeared in \cite{Hikida:2016cla}, so we have to analyze them in some way. There are several methods to evaluate these integrals, for instance, as in \cite{Diab:2016spb}. In appendix \ref{app:alternative}, we do not directly compute these integrals but rather evaluate different ones which are closely related. Using the relation between them, we deduce the $\log (x_{12}^2)$ dependence of the integrals as in \eqref{I7log}, \eqref{I8log}, and \eqref{I9log}.

From the above arguments, we conclude that the two point functions of higher spin current are corrected as 
\begin{align}
\label{jjf}
&\langle J^{\epsilon ,\pm}_s (x_1) J^{\epsilon ,\pm}_s (x_2) \rangle_\kappa
= \langle J^{\epsilon ,\pm}_s (x_1) J^{\epsilon ,\pm}_s (x_2) \rangle_0 
+ \left( \frac{16}{N} \right)^2 \frac{{\tilde{\kappa}}^2}{1 + {\tilde{\kappa}}^2 } 
\left[ \frac{1}{2C_s} \tilde I_1 + \frac{1}{2 \tilde C_s} \tilde  I_3 \right] \\
&  + \left( \frac{16}{N} \right)^2 \left( \frac{{\tilde{\kappa}}}{1 + {\tilde{\kappa}}^2 } \right)^2 
\left[ \frac{1}{2C_s} \tilde I_2 + \frac{1}{2 \tilde C_s}  \tilde I_4 \right] 
\pm  \left( \frac{16}{N} \right)^2 \left[   \frac{{\tilde{\kappa}}^2}{1 + {\tilde{\kappa}}^2 } \frac{s}{2 C_s}  \tilde I_5 + \left( \frac{{\tilde{\kappa}}}{1 + {\tilde{\kappa}}^2 } \right)^2 \frac{s}{2 C_s}  \tilde I_6 \right]  \nonumber \\
&+ \left( \frac{16}{N} \right)^4 \left( \frac{{\tilde{\kappa}}^2}{1 + {\tilde{\kappa}}^2 } \right)^2 
\left[ \frac{1}{2C_s} \tilde I_{7} + \frac{1}{2 \tilde C_s} \tilde  I_{8} 
\pm    \frac{s}{2 C_s} 
\tilde I_{9} \right] \, .\nonumber  
\end{align}
Using the $\log (x_{12}^2)$ dependence of integrals $\tilde I_a$ $(a=1, \ldots ,9)$ obtained in appendix \ref{app:fermionic} and appendix \ref{app:alternative}, we can read off the anomalous dimensions as
\begin{align}
\gamma_s^\pm  =  & \frac{16}{3 N \pi^2}  \frac{{\tilde{\kappa}}^2 }{1 + {\tilde{\kappa}}^2 } 
-  \frac{16}{ N \pi^2}  \left( \frac{{\tilde{\kappa}}}{1 + {\tilde{\kappa}}^2 } \right)^2 \frac{1}{(2s-1)(2s+1)} \nonumber \\
& \mp \left[ (-1)^s \frac{{\tilde{\kappa}}^2}{1 + {\tilde{\kappa}}^2 }+  \left( \frac{{\tilde{\kappa}}}{1 + {\tilde{\kappa}}^2 } \right)^2 \right]\frac{16}{ N \pi^2}   \frac{s}{(2s-1)(2s+1)} \label{anomalousf} \\
&- \left( \frac{{\tilde{\kappa}}^2}{1 + {\tilde{\kappa}}^2 } \right)^2\frac{16}{\pi^2 N}
\frac{s \pm 1}{(2s-1)(2s+1)} \, . \nonumber
\end{align}
This is another main result of this paper along with \eqref{anomalousb}.
We can see $\gamma_1^- = 0$ for any $\tilde \kappa$ but $\gamma_1^+ \neq 0$. This means that only a diagonal global symmetry in $\text{U}(1) \otimes \text{U}(1)$ is left unbroken. 
We can also show that $\gamma_2^+ = 0$ for even finite $\tilde \kappa$, which is consistent with the unbroken conformal symmetry.

\section{Bulk interpretations}
\label{Bulk}

In the previous sections, we have computed the anomalous dimensions of higher spin currents mainly 
in the conformal perturbation theory by following the previous works  \cite{Creutzig:2015hta,Hikida:2016wqj,Hikida:2016cla}.
% see also \cite{Hartman:2006dy,Giombi:2011ya}.
The main motivation to use the method is that  the CFT computation is closely related to the evaluation of bulk Witten diagram for one-loop corrections to the masses of dual higher spin fields as in figure \ref{2pt}. A crucial point is to rewrite bulk-to-bulk propagators with modified boundary conditions in terms of the product of bulk-to-boundary propagators and boundary two point functions \cite{Hartman:2006dy,Giombi:2011ya}. Using this, the one-loop corrections can be computed by using only tree Witten diagrams as in figure \ref{2ptmod}.
%In \cite{Creutzig:2015hta}, this way of rewriting was done for bulk scalar fields by using the embedding formulation. In the supersymmetric case, it is desired to obtain the similar result for bulk spinor fields as well, and for the purpose it seems convenient to work with the intrinsic coordinates. 
In the next subsection, we reproduce the result in \cite{Creutzig:2015hta} by using the intrinsic coordinates instead of embedding ones as was done in \cite{Hartman:2006dy,Giombi:2011ya}.
In subsection \ref{Wittens}, we examine the bulk spinor propagators by applying the method.

\subsection{Bulk scalar propagators}
\label{Wittenb}

We work on $d+1$ dimensional Euclidean AdS space, and use the Poincar\'{e} coordinates as
\begin{align}
ds^2 = \frac{1}{z_0^2 } (d z_0^2 + d\vec{z}^2) \, .
\label{Poincare}
\end{align}
Here $z_0$ is the radial coordinate and $\vec{z}$ are the coordinates parallel to the boundary, where the boundary is located at $z_0 = 0$. 
We consider two bulk scalar fields $\Phi^{(1)}$, $\Phi^{(2)}$ with the same mass but with alternative boundary conditions. 
They are dual to boundary scalar operators $\mathcal{O}_\Delta$ and $\mathcal{O}_{d- \Delta}$ with the scaling dimensions $\Delta$ and $d - \Delta$, respectively. 
In the current case, $d=3$ and $\Delta = 1$ or $2$.
The bulk-to-boundary propagators are given as \cite{Witten:1998qj,Freedman:1998tz}
\begin{align}
K_\Delta (z , \vec{x}) = \frac{\Gamma (\Delta) }{ \pi^{d/2} \Gamma (\Delta - d/2)}
\left( \frac{z_0}{z_0^2 + (\vec{z} - \vec{x})^2}\right)^\Delta \, .
\label{Witten}
\end{align}
This leads to the normalization corresponding to the standard kinetic term of dual bulk scalars as
\begin{align}
B_\Delta (\vec{x} , \vec{y}) \equiv \langle   \mathcal{O}_\Delta (\vec{x}) \mathcal{O}_\Delta ( \vec{y} ) \rangle = \frac{(2 \Delta - d) \Gamma(\Delta)}{\pi^{d/2} \Gamma(\Delta - d/2)} \frac{1}{ |\vec{x} - \vec{y}|^{2\Delta}} \, .
\end{align}
Compared with the two point functions in \eqref{2ptscalars}, the relations are
\begin{align}
\mathcal{O}_1 = \sqrt{\frac{8}{N}} \mathcal{O} \, , \quad
\mathcal{O}_2 =  \sqrt{\frac{8}{N}} \tilde{\mathcal{O}} \, . 
\end{align}
We denote $G_\Delta (z,w)$ as the bulk-to-bulk propagator, and we will need the following expression as (see, e.g., \cite{Kawano:1999au,Hartman:2006dy})
\begin{align}
\label{b2bpro}
&G_\Delta (z , w) = \int \frac{d^d k}{(2 \pi)^d} (z_0 w_0)^{d/2} e^{ i \vec{k} \cdot (\vec{z} - \vec{w})}  \\
& \times \left[ \theta (z_0 - w_0) K_{ \Delta - d/2} (k z_0) I_{ \Delta - d/2} (k w_0)  + \theta (w_0 - z_0) I_{ \Delta - d/2} (k z_0) K_{ \Delta - d/2} (k w_0)  \right] \, . \nonumber
\end{align}
Here $k = |\vec{k}|$ and $K_\nu (z)$ and $I_\nu (z)$ are the modified Bessel functions.
We deform the dual boundary CFT with
\begin{align}
\Delta_f S = f \int d^d x \mathcal{O}_{\Delta}(x) \mathcal{O}_{d - \Delta} (x)\, ,
\end{align}
where $f$ can be identified with $\tilde \lambda$ in \eqref{tlambda}.
As shown in \cite{Witten:2001ua}, this deformation is dual to mixing boundary conditions for $\Phi^{(1)}$, $\Phi^{(2)}$. Due to the changes of boundary condition, the
scalar propagators $\langle \Phi^{(a)} (z) \Phi^{(b)} (w) \rangle_f = G^{ab}_f (z,w)$ 
become \cite{Aharony:2005sh,Hikida:2015nfa}
\begin{align}
\label{propmod}
G^{ab}_f (z,w) = \frac{1}{1 + \tilde f^2} 
\begin{pmatrix}
G_\Delta  (z,w)+ \tilde f^2 G_{d - \Delta} (z,w) & \tilde f G_{d - \Delta}  (z,w) - \tilde f G_{ \Delta} (z,w)  \\
\tilde  f G_{d- \Delta}  (z,w) - \tilde f G_{ \Delta}  (z,w)  & G_{d - \Delta}  (z,w)+ \tilde f^2 G_{ \Delta}  (z,w)
\end{pmatrix}
\end{align}
with $\tilde f = (2 \Delta - d) f $. At the limit $\tilde f \to \infty$, the propagators for $\Phi^{(1)}$, $\Phi^{(2)}$ are exchanged, and this is consistent with the fact that the dual CFT  reduces to the sum of the critical U$(N)$ scalars and the Gross-Neveu model.

Here we would like to examine the bulk Witten diagram in figure \ref{2pt}, where the bulk-to-bulk scalar propagators with modified boundary conditions in \eqref{propmod} are used for the loop.
It is a quite difficult task to directly evaluate loop diagrams with higher spin gauge fields generically.
However, the AdS/CFT duality suggests that the modified scalar propagators should be written in terms of scalar propagators with $f=0$ but with the insertions of boundary operators. More precisely speaking, we should have the following identities as 
\begin{align}
\nonumber
&G_f^{11} (z,w)= G_\Delta (z,w) + f^2 \int d^d x_1 d^d x_2 K _\Delta (z, \vec{x}_1) B_{d- \Delta} (\vec{x}_1,\vec{x}_2) K_\Delta (w, \vec{x}_2)  \\
& \quad +f^4 \int \prod_{n=1}^{4} d^d x_n  K _\Delta (z, \vec{x}_1) B_{d- \Delta} (\vec{x}_1,\vec{x}_2) B_{\Delta} (\vec{x}_2,\vec{x}_3) B_{d- \Delta} (\vec{x}_3,\vec{x}_4) K_\Delta (w, \vec{x}_4) + \cdots \, , \label{Gf11} \\
\nonumber
&G_f^{12} (z,w)= - f \int d^d x K _\Delta (z, \vec{x})  K_{d -\Delta} (w, \vec{x})  \\
& \quad - f^3 \int \prod_{n=1}^{3} d^d x_n  K _\Delta (z, \vec{x}_1) B_{d- \Delta} (\vec{x}_1,\vec{x}_2) B_{\Delta} (\vec{x}_2,\vec{x}_3)  K_{d- \Delta} (w, \vec{x}_3) + \cdots \, . \label{Gf12}
\end{align}
The same should be true for $G_f^{22}$ and $G_f^{21}$ by exchanging $\Delta$ with $d - \Delta$.
With these identities, the loop Witten diagram in figure \ref{2pt} can be evaluated only from the product of tree Witten diagrams as in figure \ref{2ptmod}.%
\footnote{Here we have used the fact that there is no contribution to the two point function from the one-loop diagram at $f=0$.}
Moreover, we can see that the latter description is dual to the boundary computation in conformal perturbation theory.
Therefore, we can obtain the map between the bulk computation with Witten diagrams and boundary one in conformal perturbation theory, once we can confirm the identities \eqref{Gf11} and \eqref{Gf12}.
We can check that the identities \eqref{Gf11} and \eqref{Gf12} follow the two basic ingredients 
\begin{align}
& \int d^d x K_\Delta (z , \vec{x}) B_{d - \Delta} (\vec{x} , \vec{y}) 
 = (d - 2 \Delta) K_{d - \Delta} (z, \vec{x}) \, , \label{KB} \\
& \int d^d x K_\Delta (z , \vec{x}) K_{d - \Delta} (w , \vec{x})
  = (2 \Delta - d) \left[  G_\Delta (z , w) - G_{d - \Delta} (z,w) \right] \, .  \label{KK}
\end{align}
In the rest of this subsection, we shall derive them.
They were already shown as (B.21) and (5.7) in  \cite{Creutzig:2015hta} by adopting the embedding formulation. 
For the extension to bulk spinor propagators in next subsection, we re-derive the results using the intrinsic coordinates in \eqref{Poincare}. The derivation of \eqref{KK} here is the same as the one in \cite{Hartman:2006dy}, see also \cite{Giombi:2011ya}.

For our purpose, it is convenient to work in the momentum representation as
\begin{align}
&K_\Delta (z_0 , \vec{k}) = \int \frac{d^d z}{(2 \pi)^{d/2}} K_\Delta (z , 0) 
e^{- i \vec{k} \cdot \vec{z}  } = \frac{2^{1 - \Delta} z_0^{d/2}}{\pi^{d/2} \Gamma(\Delta - d/2)} k^{\Delta - d/2} K_{\Delta - d/2} (z_0 k) \, , \nonumber \\
&B_\Delta (\vec{k}) = \int \frac{d^d x}{(2 \pi)^{d/2}} B_\Delta ( \vec{x} , 0) 
e^{- i \vec{k} \cdot \vec{x} }
= \frac{2^{d/2 - 2 \Delta} (2 \Delta - d) \Gamma(d/2 - \Delta)}{\pi^{d/2} \Gamma(\Delta - d/2)} k^{2 \Delta- d} \, .
\end{align}
Using these expressions, we can show 
\begin{align}
&\int d^d x K_\Delta (z , \vec{x}) B_{d - \Delta} (\vec{x} , \vec{y}) 
= \int d^d x  \frac{d^d k_1}{(2 \pi)^{d/2}}\frac{d^d k_2}{(2 \pi)^{d/2}}
K_\Delta (z_0 , \vec{k}_1) e^{i \vec{k}_1 \cdot (\vec{z} - \vec{x}) } B_{d - \Delta}(\vec{k}_2)
 e^{ i \vec{k}_2 \cdot (\vec{x} - \vec{y}) } \nonumber \\
 &= \frac{2^{1 - 3d/2 + \Delta} z_0^{d/2} (d - 2 \Delta)}{\pi^{d} \Gamma(d/2 - \Delta)} 
 \int d^d k e^{i \vec{k} \cdot (\vec{z} - \vec{y})}
k^{d/2 - \Delta } K_{d/2 - \Delta } (z_0 k) 
 = (d - 2 \Delta) K_{d - \Delta} (z_0 , \vec{z} , \vec{y})
\end{align}
as in \eqref{KB}.
Similarly, we have  for \eqref{KK}
\begin{align}
&\int d^d x K_\Delta (z , \vec{x}) K_{d - \Delta} (w , \vec{x}) \\
&= 
 \int d^d x  \frac{d^d k_1}{(2 \pi)^{d/2}}\frac{d^d k_2}{(2 \pi)^{d/2}}
K_\Delta (z_0 , \vec{k}_1) e^{i \vec{k}_1 \cdot (\vec{z} - \vec{x}) }
K_{d - \Delta} (w_0 , \vec{k}_2) e^{i \vec{k}_2 \cdot (\vec{w} - \vec{x}) } \nonumber \\
& = \frac{2^{2 - d} (z_0 w_0)^{d/2}}{\pi^d \Gamma(\Delta - d/2) \Gamma(d/2 - \Delta)}
 \int d^d k K_{\Delta - d/2} (z_0 k) K_{\Delta -d/2} (w_0 k) e^{i \vec{k} \cdot (\vec{z} - \vec{w}) }
\, . \nonumber
\end{align}
Applying the formula $(\nu = \Delta - d/2)$
\begin{align}
K_\nu (z_0 k) =  \frac{\nu}{2} \Gamma(\nu) \Gamma(- \nu) [ I_\nu (z_0 k) -  I_{- \nu} (z_0 k ) ] \, , 
\end{align}
we find
\begin{align}
& \int d^d k K_{\nu} (z_0 k) K_{\nu} (w_0 k) = 
  \int d^d k K_{\nu} (z_0 k) K_{\nu} (w_0 k) 
[\theta (z_0 - w_0) + \theta (w_0 - z_0)] \nonumber  \\
&= \frac{\nu}{2} \Gamma(\nu) \Gamma(- \nu ) 
 \theta (z_0 - w_0) \int d^d k K_\nu (z_0 k)  (I_{\nu} (w_0 k) -  I_{- \nu} (w_0 k) ) \\
& \quad +  \frac{\nu}{2} \Gamma(\nu) \Gamma(- \nu )  \theta (z_0 - w_0) \int d^d k K_\nu (w_0 k) (I_{\nu} (z_0 k) -  I_{- \nu} (z_0 k) ) 
 \, . \nonumber
\end{align}
With the expression of bulk-to-bulk propagators in \eqref{b2bpro},
we arrive at \eqref{KK}.

\subsection{Bulk spinor propagators}
\label{Wittens}

There is another type of marginal deformation as in \eqref{defsusy}, and they should be dual to modifying boundary conditions for bulk spinor fields. Thus the anomalous dimensions of higher spin currents can be computed from one-loop Witten diagram as in figure \ref{2pt}, but now bulk spinor fields with modified boundary condition are running along the loop. In this subsection, we show that the one-loop Witten diagram can be evaluated from the product of tree Witten diagrams as in figure \ref{2ptmod} just like for the bulk scalar propagators.

We examine $d+1$ dimensional Euclidean space with the coordinate system \eqref{Poincare}. Moreover, we consider two bulk Dirac fermions $\Psi ^{(1)}, \Psi ^{(2)}$ with the same mass $m$ but alternative boundary conditions. 
The spinor operators dual to $\Psi ^{(1)}, \Psi ^{(2)}$ are given as $\mathcal{K}_{\Delta_+},\mathcal{K}_{\Delta_-}$, whose conformal dimensions are $\Delta_\pm = d/2 \pm m$. 
For application to the current case, we should set $d=3$, $m=0$, and consider only half of the fermions. 

Spinor propagators in AdS$_{d+1}$ have been investigated in \cite{Henningson:1998cd,Mueck:1998iz,Kawano:1999au}, and we follow their analysis. 
We introduce the Gamma matrices $\Gamma^a$ ($a=0,1,\cdots,d$) satisfying 
$\{ \Gamma^a , \Gamma^b \} = 2 \delta^{ab}$.
Defining 
\begin{align}
U(z , \vec{x} ) = z_0^{1/2} \Gamma^0 + z_0^{-1/2} \vec{\Gamma} \cdot
 (\vec{z} - \vec{x}) \, , 
\end{align}
the bulk-to-boundary propagators are given as 
\begin{align}
\Sigma_{\Delta_\pm} (z , \vec{x}) = U(z , \vec{x}) K_{\Delta_\pm + 1/2} (z , \vec{x}) \mathcal{P}_\mp \, ,  \quad
\bar{\Sigma}_{\Delta_\pm} (z , \vec{x}) = \pm \mathcal{P}_\pm K_{\Delta_\pm + 1/2} (z , \vec{x})  U(z , \vec{x}) \, , 
\label{bulk2bdryf}
\end{align}
where
\begin{align}
\mathcal{P}_\pm = \frac{1}{2} (1 \pm \Gamma^0) \, .
\end{align}
The boundary two point functions are
\begin{align}
F_{\Delta_{\pm}} (\vec{x} , \vec{y}) \equiv \langle \mathcal{K}_{\Delta_\pm } (\vec{x}) 
\bar{\mathcal{K}}_{\Delta_\pm } (\vec{y}) \rangle 
 = \frac{\Gamma (\Delta_\pm + 1/2)}{\pi^{d/2} \Gamma ( 1/2 \pm m)} 
    \frac{\vec{\Gamma} \cdot (\vec{x} - \vec{y})}{|\vec{x} - \vec{y}|^{2 \Delta_\pm + 1}} 
    \mathcal{P}_\mp \, .
    \label{bdry2bdryf}
\end{align}
Compared with \eqref{2ptsfermions},
the normalization of spinor operators differs from $\mathcal{K}$ by the factor $\sqrt{16/N}$.

As in \cite{Kawano:1999au}, the bulk-to-bulk propagators should satisfy
\begin{align}
(\slashed{D} - m) S^m_\pm (z,w) = S^m_\pm (z , w) (- \overleftarrow{\slashed{D}} - m) = \frac{1}{\sqrt{g}} \delta^{(d+1)} (z - w) \, ,
\end{align}
and furthermore regularity and boundary condition. 
Solutions were obtained as 
\begin{align}
S_\pm^m (z , w ) =& \mp \int \frac{d^d k}{(2 \pi)^d} k e^{i \vec{k} \cdot (\vec{z} - \vec{w})}
\Biggl[ \theta (z_0 - w_0) \phi_{\pm,m}^{(K) } (z_0 , \vec{k}) \mathcal{P}_\mp 
\bar{\phi}^{(I)}_{\mp,m} (w_0 , - \vec{k})  \\
& \qquad - \theta (w_0 - z_0) \phi_{\pm,m}^{(I) } (z_0 , \vec{k}) \mathcal{P}_\mp 
\bar{\phi}^{(K)}_{\mp,m} (w_0 , - \vec{k})  \Biggr] \nonumber \, .
\end{align}
Here we have defined
\begin{align}
&\phi^{(K)}_{\pm , m} (z_0 , \vec{k}) =  z_0^{\frac{d+1}{2}} \left[ K_{m \pm 1/2} (k z_0) \pm i \frac{\slashed{k}}{k} K_{m \mp 1/2} (k z_0) \right]  \, , \\
&\phi^{(I)}_{ \pm , m } (z_0 , \vec{k}) = z_0^{\frac{d+1}{2}} \left[ I_{m \pm 1/2} (k z_0) \mp i \frac{\slashed{k}}{k} I_{m \mp 1/2} (k z_0) \right] \, , 
\end{align}
and
\begin{align}
&\bar{\phi}^{(K)}_{ \mp , m} (z_0 , \vec{k}) =   z_0^{\frac{d+1}{2}} \left[ K_{m \mp 1/2} (k z_0) \pm i \frac{\slashed{k}}{k} K_{m \pm 1/2} (k z_0) \right]  =  \phi^{(K)}_{\mp , m} (z_0 , - \vec{k})\, , \\
&\bar{\phi}^{(I)}_{ \mp , m } (z_0 , \vec{k}) =   z_0^{\frac{d+1}{2}} \left[ I_{m \mp 1/2} (k z_0) \mp i \frac{\slashed{k}}{k} I_{m \pm 1/2} (k z_0) \right]  
= \phi^{(I)}_{ \mp , m } (z_0 , - \vec{k}) \, .
\end{align}
Since we have
\begin{align}
(\slashed{D} - m) S^m_\pm (z,w) = \left( z_0 \Gamma^a \partial_a - \frac{d}{2} \Gamma_0 - m \right) S^m_\pm (z,w)   \, ,
\end{align}
the solution with $-m$ can be obtained from $S^m_\pm (z,w)$ by replacing $\Gamma^0 $ by $- \Gamma^0$ and $(z,w)$ by $(-z,-w)$. This implies that 
\begin{align}
S^{-m}_\pm (z,w) = - S^m_\mp (-z , -w)  \, ,
\label{-m2m}
\end{align}
which will be used for later analysis.

We deform the dual boundary CFT by
\begin{align}
\Delta _h S = h \int d^d x \left[  \bar{\mathcal{K}}_{\Delta_+}  (x) \mathcal{K}_{\Delta_-} (x) + \bar{\mathcal{K}}_{\Delta_-} (x)\mathcal{K}_{\Delta_+} (x) \right] \, .
\end{align}
Here we have two terms since now we are working with two Dirac fermions. The deformation parameter $h$ can be identified with $\tilde \kappa$ in \eqref{tkappa}.
The deformation is dual to mixing the boundary conditions for $\Psi^{(1)}$, $\Psi^{(2)}$,
and the modified spinor propagators $\langle \Psi^{(a)} (z) \bar{\Psi}^{(b)} (w) \rangle = S^{ab}_f (z,w)$ are \cite{Hikida:2015nfa}
\begin{align}
S^{ab}_h (z,w) = \frac{1}{1 + h^2} 
\begin{pmatrix}
S^m_+ (z,w)+  h^2 S^m_- (z,w) & h S^m_-  (z,w) - h S^m_+ (z,w)  \\
h S^m_- (z,w) - h S^m_+ (z,w)  & S^m_-  (z,w)+ h^2 S^m_+  (z,w)
\end{pmatrix} \, . 
\label{msp}
\end{align}
We can see that  the propagators for $\Psi^{(1)}$, $\Psi^{(2)}$ are exchanged at the limit $h \to \infty$.
Below we shall show that 
\begin{align}
&\int d^d x \Sigma_{\Delta_\pm} (z , \vec{x}) F_{\Delta_\mp} (\vec{x} , \vec{y}) 
= \pm \Sigma_{\Delta_\mp} (z , \vec{y}) \, , \label{SF} \\
&\int d^d x F_{\Delta_\mp} (\vec{x} , \vec{y}) \bar \Sigma_{\Delta_\pm} (z , \vec{y}) 
= \pm \bar \Sigma_{\Delta_\mp} (z , \vec{x}) \, , \label{SF2} \\
&\int d^d x \Sigma_{\Delta_\pm} (z , \vec{x} )  \bar{\Sigma}_{\Delta_\mp} (w , \vec{x} ) 
= \pm (S^m_\pm (z,w) - S^m_{\mp} (z,w) )\, . \label{SS} 
\end{align}
Using these identities, we can rewrite the modified spinor propagators \eqref{msp} as those with $h=0$ but with the insertions of boundary operators precisely as for the bosonic case, see \eqref{Gf11} and \eqref{Gf12}.

In order to proof the identities \eqref{SF}, \eqref{SF2} and \eqref{SS}, it is convenient to express the propagators \eqref{bulk2bdryf} and \eqref{bdry2bdryf} in the momentum representation as
\begin{align}
&\Sigma_{\Delta_{\pm}} (z_0 , \vec{k} ) 
= \int \frac{d^d z}{(2 \pi)^{d/2}} \Sigma_{\Delta_{\pm}} (z , 0) e^{ - i \vec{k} \cdot \vec{z} } =  \mp \frac{2^{1/2 - \Delta_\pm }}{\pi^{d/2} \Gamma (1/2 \pm m  )} k^{\pm m  + 1/2 } \phi^{(K)}_{\pm ,m} (z_0 , \vec{k}) \mathcal{P}_\mp \, ,  \\
& \bar{\Sigma}_{\Delta_{\pm}} (z_0 ,  \vec{k} ) 
 = \int \frac{d^d z}{(2 \pi)^{d/2}} \bar{\Sigma}_{\Delta_{\pm}} (z , 0) e^{ - i \vec{k} \cdot \vec{z} } =  \mathcal{P}_{\pm} \frac{2^{1/2 - \Delta _\pm }}{\pi^{d/2} \Gamma (1/2 \pm m )} k^{\pm  m + 1/2 } \bar{\phi}^{(K)}_{\pm , m} (z_0 , \vec{k}) \, ,
\end{align}
and
\begin{align}
F_{\Delta_{\pm}}( \vec{k})  = \int \frac{d^d z}{(2 \pi)^{d/2}} F_{\Delta_\pm} (\vec{x} , 0) e^{ - i \vec{k} \cdot \vec{x} } 
=  - \frac{\Gamma (1/2 \mp m )}{2^{2 \Delta_\pm - d/2} \pi^{d/2} \Gamma(1/2 \pm m)} i \slashed{k} k^{2 \Delta_\pm - 1 - d}\mathcal{P}_\mp \, .
\end{align}
Then we can show
\begin{align}
\int d^d x \Sigma_{\Delta_\pm} (z , \vec{x}) F_{\Delta_\mp} (\vec{x} , \vec{y}) 
 = \int d^d k \Sigma_{\Delta_\pm} (z , \vec{k}) F_{\Delta_\mp} (- \vec{k}) e^{i \vec{k} \cdot (\vec{z} - \vec{y})}
 = \pm \Sigma_{\Delta_\mp} (z , \vec{y}) 
\end{align}
for \eqref{SF} and similarly for \eqref{SF2}.
For \eqref{SS} we rewrite
\begin{align}
\nonumber
&\int d^d x \Sigma_{\Delta_\pm} (z , \vec{x} )  \bar{\Sigma}_{\Delta_\mp} (w , \vec{x} ) 
= \int d^d k e^{i \vec{k} \cdot (\vec{z} - \vec{w})} \Sigma_{\Delta_\pm} (z_0 , \vec{k} )  \bar{\Sigma}_{\Delta_\mp} (w_0 , - \vec{k} )  \\
&= \mp \frac{2^{1 - d}}{\pi^d \Gamma(1/2 \pm m ) \Gamma (1/2 \mp m )}  \int d^d k e^{i \vec{k} \cdot (\vec{z} - \vec{w})} k \phi^{(K)}_{\pm , m} (z_0 , \vec{k} ) \mathcal{P}_\mp \bar{\phi}_{\mp ,m}^{(K)} (w_0 , - \vec{k}) \, .
\end{align}
Using 
\begin{align}
 \phi^{(K)}_{\pm , m} (z_0 , \vec{k}) =  \phi^{(K)}_{\mp ,- m} (z_0 , - \vec{k}) \, , 
\end{align}
and
\begin{align}
\phi^{(K)}_{\pm ,m} (z_0 , \vec{k}) = \mp \frac{1}{2} \Gamma(1/2 \pm m ) \Gamma (1/2 \mp m )  \left[ \phi^{(I)}_{\pm , m} (z_0 , \vec{k}) - \phi^{(I)}_{\mp , -m} (z_0 , - \vec{k})\right] \, ,
\end{align}
we find \eqref{SS} with the help of \eqref{-m2m}. 

\section{Conclusion and discussions}
\label{sec:conclusion}

In this paper, we have studied 3d supersymmetric U$(N)$ model, which is supposed to be dual to 4d supersymmetric Vasiliev theory \cite{Leigh:2003gk,Sezgin:2003pt}.
The model admits two types of marginal deformation as in \eqref{def} and \eqref{defsusy}, and the deformations should be dual to modifying boundary conditions for dual scalars and spinors, respectively \cite{Witten:2001ua}.
There are two main results of this paper.
One of them is on the anomalous dimensions of higher spin currents in the deformed models.
Using basically conformal perturbation theory, we obtained them to the leading order in $1/N$ but to all orders in deformation parameters as in \eqref{anomalousb} and \eqref{anomalousf}. 
For higher order corrections in $\kappa$ in the deformed theory by \eqref{defsusy}, we utilized another technique as used in \cite{Muta:1976js}.
The other is on the dual higher spin interpretation of the computations in conformal perturbation theory, which can be done by rewriting one-loop Witten diagram as in figure \ref{2pt} to the product of tree Witten diagrams as in figure \ref{2ptmod}. We have derived the ingredients essential for the rewritening in \eqref{KB}, \eqref{KK} for bulk scalars and in \eqref{SF}, \eqref{SF2}, \eqref{SS} for bulk spinors. This was already done in \cite{Creutzig:2015hta} for bulk scalars (see \cite{Hartman:2006dy,Giombi:2011ya} for \eqref{KK}) but it was new for bulk spinors.

We examined the 3d supersymmetric model as a simple model which has marginal deformations and higher spin holographic dual. We would like to relate deforming the CFT marginally and turning on the string tension. In order to reveal more precise relations, however, we need to extend the analysis to more complicated systems. For instance, a concrete relation (named as ABJ triality) was proposed in \cite{Chang:2012kt} through 3d ABJ(M) theory in \cite{Aharony:2008ug,Aharony:2008gk}.
For application to the ABJ triality, we have to work with coupling to Chern-Simons gauge fields as in \cite{Giombi:2011kc,Aharony:2011jz,Maldacena:2012sf}. 
See \cite{Giombi:2016zwa} for a recent work.
There are also lower dimensional proposals with 2d CFTs which have $\mathcal{N}=4$ supersymmetry in \cite{Gaberdiel:2013vva,Gaberdiel:2014cha} and $\mathcal{N}=3$ supersymmetry in \cite{Creutzig:2013tja,Creutzig:2014ula,Hikida:2015nfa}. 
Symmetry breaking for higher spin gauge theory on AdS$_3$ has been studied in \cite{Hikida:2015nfa,Creutzig:2015hta,Gaberdiel:2015uca,Gwak:2015jdo}, but more detailed analysis would be required to say more concretely.

There are other open problems as follows;
We have evaluated the log$(x_{12}^2)$ dependence of integrals $\tilde I_{7}$, $\tilde I_{8}$, $\tilde I_{9}$ in appendix \ref{app:alternative} with an indirect method. However, it is desired to evaluate them explicitly, see also footnote \ref{muta}.
A drawback of our method is that we cannot identify the Nambu-Goldstone modes arising due to symmetry breaking as mentioned in \cite{Hikida:2016wqj,Hikida:2016cla}.  A direct way to evaluate the loop corrections to the bulk higher spin propagators explicitly, but it looks a quite hard task. 
Along with the difficulty for evaluating loops, there is an additional problem on our limited understanding of the dual Vasiliev theory, where only classical equations of motion are available presently.
An indirect way might be given by group theoretic analysis as was done for the 4d Vasiliev theory dual to 3d critical O$(N)$ model in \cite{Girardello:2002pp}. It would be also possible by writing the divergence of dual higher spin current in terms of products of higher spin currents, see, e.g., \cite{Maldacena:2012sf,Skvortsov:2015pea,Giombi:2016hkj,Giombi:2016zwa}.
The 3d supersymmetric U$(N)$ model itself is also worth studying furthermore.
The deformations in \eqref{def} and \eqref{defsusy} are exactly marginal to the $1/N$ order, but it is not clear what happens at the next order. Even if we could find non-trivial fixed lines (or points), it is notorious to be a hard problem to go beyond the leading order in $1/N$. 
Hopefully, this issue would be tractable along the line in, e.g., \cite{Vasiliev:1975mq,Vasiliev:1993ux,Derkachov:1997ch,Manashov:2016uam}.
It is also important to extend our analysis to, say, the theory with both bosons and fermions in generic $d \neq 3$ as in \cite{Gunaydin:2016amv,Giombi:2016pvg}. In the case, the deformations \eqref{def} and \eqref{defsusy} are not marginal anymore, so we may examine the fixed points of RG flow induced by these deformations.

\subsection*{Acknowledgements}

The work of YH is supported by JSPS KAKENHI Grant Number 24740170, 16H02182. 

\appendix

\section{Evaluations of integral}

In this appendix, we examine the integrals $I_a$ $(a=1,\ldots,5)$
and $\tilde I_a$ $(a=1,\ldots,6)$ defined in the main context. We denote the $\log (x_{12}^2)$ dependence of these integrals 
as $I_a^{\log}$ and $\tilde I_a^{\log}$, and we shall obtain them explicitly by borrowing the results in \cite{Hikida:2016wqj,Hikida:2016cla}.

\subsection{Deformation with bosonic operators}
\label{app:bosonic}

In the case with \eqref{def}, the integrals are almost the identical to those evaluated in \cite{Hikida:2016wqj,Hikida:2016cla}, so we can directly use the results there.
The integral $I_1$ in \eqref{I1} is almost the same as the one in section 3.2 of \cite{Hikida:2016cla} with $d=3$, once we replace the correlator
\begin{align}
\langle \tilde{\mathcal{O}}(x_1) \tilde{\mathcal{O}} (x_2) \rangle _0 = \frac{2 N C_\phi^2}{(x_{12}^2)^{2}}
\end{align}
by the propagator of auxiliary field $\sigma$. Therefore, we may shift the exponent as 
$(x_{12}^2)^{-2} \to (x_{12}^2)^{-2 + \Delta} $.
Then the reside at $\Delta=0$ should be the factor in front of $\log (x_{12}^2)$. 
Since we are working with complex scalars, we need to multiply some factors as discussed in section 5.2 of \cite{Hikida:2016cla}. 
Moreover, we have to replace the overall factor $C_\sigma$ for the auxiliary field propagator by $2NC_\phi^2$.
We then find
\begin{align}
I_1^{\log} = - \frac{N}{12 \pi^2} \frac{(s-1)(s+1)}{(2s - 1)(2s+1)} D_0^s \log(x_{12}^2) \, .
\end{align}
We can evaluate $I_2$ in \eqref{I2} as for $I_1$.
Replacing
\begin{align}
\langle \mathcal{O}(x_1) \mathcal{O} (x_2) \rangle _0 =  \frac{N C_\phi^2}{x_{12}^2} 
\end{align}
by the propagator of auxiliary field $\tilde \sigma$, the computation of $I_2$ is almost the same as that of $I_1$ in subsection 4.2 of \cite{Hikida:2016cla}. Thus we shift as $(x_{12}^2)^{-1} \to (x_{12}^2)^{-1 + \Delta}$
and pick up the $1/\Delta$-pole term. Replacing  the overall factor $C_{\tilde \sigma}$ for the auxiliary field propagator by $N C_\phi^2$, we find
\begin{align}
I_2 ^{\log}= - \frac{N}{12 \pi^2} \frac{(s-1)(s+1)}{(2s - 1)(2s+1)} \tilde D_0^s \log (x_{12}^2) \, .
\end{align}

The integral $I_3$ in \eqref{I3} vanishes for $s$ odd, so we set $s$ even here.
Using the three point functions obtained in \cite{Hikida:2016cla}, the integral $I_3$ can be written as
\begin{align}
I_3 = 8 N^2 C_\phi^6  \sum_{k,l=0}^s a_k a_l B_{k,l} 
\end{align}
with
\begin{align}
B_{k,l} = \int d^3 x_3 d^3 x_4 \frac{1}{(x_{34}^2)^2}
\left[ \hat \partial_1^{s-k} \frac{1}{(x_{41}^2)^{1/2} }\right] 
\left[ \hat \partial_1^{k} \frac{1}{(x_{31}^2)^{1/2} }\right] 
\left[ \hat \partial_2^{l} \frac{1}{(x_{42}^2)^{1/2} }\right] 
\left[ \hat \partial_2^{s-l} \frac{1}{(x_{32}^2)^{1/2} }\right] \, .
\label{Bkl}
\end{align}
The same integral arises from the O$(N)$ scalars as $I_1^{(2)}$ in subsection 3.2.2 of  \cite{Hikida:2016cla},
so we use the regularization adopted there.
We evaluate the integral with $(x_{34}^2)^{-2}$ replaced by $(x_{34}^2)^{-2  + \Delta}$.
Borrowing the result in \cite{Hikida:2016cla}  as
\begin{align}
\sum_{k,l=0}^s a_k a_l B_{k,l} = \frac{1}{2N C_\phi^4} \frac{1}{(2s-1)(2s+1)} D_0^s  \frac{1}{\Delta}
+ \mathcal{O} (\Delta^0) \, , 
\end{align}
we find
\begin{align}
I_3 ^{\log}  =  \mathcal{P}_s \frac{N}{4 \pi^2} \frac{1}{(2s-1)(2s+1)} D_0^s \log (x_{12}^2) \, .
\end{align}

For the integral $I_4$ in \eqref{I4}, we adopt the regularization used for $I_1$. 
Then the integral is the same as the one computed in section 3.3 of \cite{Hikida:2016cla}. 
Since we are working with U$(N)$ global symmetry, we need to multiply each three point function by $1/4$. Moreover, we should replace $C_\sigma$ by $2N C_\phi$.
Thus we find
\begin{align}
I_4^{\log}  = \mathcal{P}_s \frac{N^3}{2^{9} \pi^2} \frac{s}{(2s-1)(2s+1)} D_0^s \log (x_{12}^2) \, .
\end{align}
For the other integral  $I_5$ in \eqref{I5}, we adopt the regularization used for $I_2$. 
Then the integral is the same as the one computed in subsection  4.3. 
Replacing $C_{\tilde \sigma}$ by $N C_\phi$, we obtain
\begin{align}
I_5^{\log}  =  \mathcal{P}_s \frac{N^3}{2^{9} \pi^2} \frac{s}{(2s-1)(2s+1)}\tilde D_0^s \log (x_{12}^2) \, .
\end{align}
Here we should remark that the contributions to the $\log (x_{12}^2)$ terms are not the $1/\Delta$-pole terms but twice of them. See \cite{Diab:2016spb} for more details, for instance. 

\subsection{Deformation with fermionic operators}
\label{app:fermionic}

We move to examine the integrals appeared in section \ref{sec:fermion}, where we have deformed the theory by \eqref{defsusy}.
The integral $\tilde I_1$ in \eqref{tI1} is computed as
\begin{align}
\tilde I_1
&=  4  N^2 C_\phi^6 
\int d^3 x_3 d^3 x_4 
\sum_{k,l=0}^{s}a_k  a_l \left[ \hat \partial^{k}_1 \frac{1}{(x_{14}^2)^{1/2}} \right]
\left[  \hat \partial^{s-k}_1 \hat \partial^l_2 \frac{1}{(x_{12}^2)^{1/2}} \right]
\left[ \hat \partial^{s-l}_2 \frac{1}{(x_{23}^2)^{1/2}}  \right]
\frac{1}{(x_{34}^2)^{5/2} } \, .
\nonumber 
\end{align}
The integral is essentially the same as $I_1^{(1)}$ computed in subsection 3.2.1 of \cite{Hikida:2016cla}, so we evaluate the integral in the same way. Namely, we replace 
$(x_{34}^2)^{-5/2}$ by $(x_{34}^2)^{-5/2 +\Delta}$ and pick up the $1/\Delta$-pole term. 
The $\log (x_{12}^2)$ dependence can be read off as
\begin{align}
\tilde I_1^{\log } =  - \frac{N}{3\cdot 2^4 \pi^2} D_0^s \log (x_{12}^2) \, .
\end{align}
The integral $\tilde I_2$ in \eqref{tI2} can be written as
\begin{align}
\tilde I_2
= 2 N^2 C_\phi^6 
\sum_{k,l=0}^{s} a_k a_l  B_{k,l} \, ,
\end{align}
where $B_{k,l}$ was defined in \eqref{Bkl}.
As for $I_3$, we find
\begin{align}
\tilde I_2 ^{\log}= \frac{N}{ 2^4 \pi^2}\frac{1}{(2s-1)(2s+1)} D_0^s \log (x_{12}^2)\, .
\label{tI2log}
\end{align}

The integral $\tilde I_3$ in \eqref{tI3} becomes
\begin{align}
& \tilde I_3
=  \frac{2}{3} N^2 C_\phi^6 
\int d^3 x_3 d^3 x_4  \\
& \times
\sum_{k,l=0}^{s-1}\tilde a_k \tilde a_l  \text{tr} \left[
\hat \partial^k_1 \slashed{\partial}_3 \frac{1}{(x_{13}^2)^{1/2}} \right]
\hat \gamma \left[\hat \partial^{s-k-1}_1 \hat \partial^l_2 \slashed{\partial}_1 \frac{1}{(x_{12}^2)^{1/2}}  \right] \left[ \hat \gamma \hat \partial^{s-l-1}_2 \slashed{\partial}_2 \frac{1}{(x_{24}^2)^{1/2}} \right] \left[ \slashed{\partial}_3 \frac{1}{(x_{34}^2)^{3/2}} 
\right] .
\nonumber 
\end{align}
The integral is almost the same as $I_1^{(1)}$ in subsection 4.2.1 of \cite{Hikida:2016cla}, so we shift $(x_{34}^2)^{-3/2}$ as $(x_{34}^2)^{-3/2 + \Delta}$.
From the $1/\Delta$-pole term, we find
\begin{align}
\tilde I_3 ^{\log} = - \frac{N}{3\cdot 2^4 \pi^2} \tilde D_0^s \log (x_{12}^2) \, .
\end{align}
Similarly, we obtain
\begin{align}
&\tilde I_4 
= -N^2 C_\phi^6 \sum_{k,l = 0}^{s-1} \tilde a_k \tilde a_l  \int d^3 x_3 d^3 x_4 
\\ &   \times
\text{tr} \left\{
\left[\hat \partial^k _1 \slashed{\partial}_3 \frac{1}{(x_{13}^2)^{1/2}}  \right] \hat \gamma
\left[ \hat \partial_1^{s-1-k} \slashed{\partial}_1  \frac{1}{(x_{14}^2)^{1/2}} \right] 
\left[ \hat \partial^{l}_2 \slashed{\partial}_4 \frac{1}{(x_{24}^2)^{1/2}}   \right]
\hat \gamma \left[
\hat \partial^{s-1- l}_2 \slashed{\partial}_2 \frac{1}{(x_{23}^2)^\delta} \right] \frac{1}{x_{34}^2} \right\} \, . \nonumber
\end{align}
The integral is almost the same as $I_1^{(2)}$ in subsection 4.2.2 of \cite{Hikida:2016cla}, so we change $(x_{34}^{2})^{-1}$ to $(x_{34}^{2})^{-1 + \Delta}$. From the $1/\Delta$-pole term, we have
\begin{align}
\tilde I_4 ^{\log}= \frac{N}{ 2^4 \pi^2}\frac{1}{(2s-1)(2s+1)} \tilde D_0^s \log (x_{12}^2) \, .
\end{align}

The integrals  $\tilde I_5$ in \eqref{tI5} and $\tilde I_6$ in \eqref{tI6} are of the similar form as
\begin{align}
&\tilde I_5
=  (-1)^s \tilde I_6 = \frac{1}{2} N^2 C_\phi^6 \sum_{k=0}^{s} \sum_{l=0}^{s-1} a_k \tilde  a_l \int d^3 x_3 d^3 x_4 \\
& \times  \text{tr} \left\{\left[
\hat \partial^k_1 \frac{1}{(x_{14}^2)^{1/2}} \right] \left[ \hat \partial^{s-k}_1 \frac{1}{(x_{13}^2)^{1/2}} \right] \left[
\hat \partial^l_2 \slashed{\partial}_3 \frac{1}{(x_{23}^2)^{1/2}} \right] 
\hat \gamma \left[ \hat \partial^{s-l-1}_2 \slashed{\partial}_2 \frac{1}{(x_{24}^2)^{1/2}} \right] \left[
\slashed{\partial}_3 \frac{1}{x_{34}^2} \right]  \right\} \, .
\nonumber 
\end{align}
Using the relation (see (4.41) and (4.42) of \cite{Hikida:2016cla}) 
\begin{align}
\sum_{l=0}^{s-1} \tilde  a_l \text{tr} \left[
\hat \partial^l_2 \slashed{\partial}_2 \frac{1}{(x_{24}^2)^{1/2}} \right] 
\hat \gamma \left[ \hat \partial^{s-l-1}_2 
\slashed{\partial}_2 \frac{1}{(x_{23}^2)^{1/2}} \right]
\left[\slashed{\partial}_3 \frac{1}{x_{34}^2} \right]  \nonumber \\
= \frac{8}{(x_{23}^2)^{3/2}} \sum_{l=0}^{s}a_l \left[
\hat \partial^l_2 \frac{1}{(x_{24}^2)^{1/2}} \right] \left[ \hat \partial^{s-l}_2 \frac{1}{(x_{23}^2)^{1/2}} \right] \, ,
\nonumber 
\end{align}
we can rewrite as
\begin{align}
\tilde I_5
= & (-1)^s \tilde I_6 = (-1)^s 4 N^2 C_\phi^6  \sum_{k,l=0}^{s}  a_k a_l B_{k,l} 
\end{align}
with $B_{k,l}$ in \eqref{Bkl}.
The integral $\tilde I_6$ is twice of $\tilde I_2$ evaluated as in \eqref{tI2log},
which leads to
\begin{align}
\tilde I_5 ^{\log} = (-1)^s \tilde I_6 ^{\log} = (-1)^s \frac{N}{ 8 \pi^2}\frac{1}{(2s-1)(2s+1)} D_0^s \log (x_{12}^2) \, .
\end{align}

\section{Another method for anomalous dimensions}
\label{app:alternative}

In the previous appendix, we have obtained  the $\log (x_{12}^2)$ dependence of integrals 
$I_a$ $(a=1,\ldots,5)$ and $\tilde I_a$ $(a=1,\ldots,6)$.
Therefore, only the integrals $\tilde I_7,\tilde I_8,\tilde I_9$ in \eqref{tI7}, \eqref{tI8}, \eqref{tI9} are left.
In principal, it is straightforward to evaluate the integrals as in \cite{Diab:2016spb}, but it is quite tedious. In this section, we compute the $\log (x_{12}^2)$ dependence by developing more simple but indirect method.

For explanation, let us focus on the case of the Gross-Neveu model. In case with our main method, we compute $\langle \tilde J^\epsilon_s (p) \tilde J^\epsilon_s (-p) \rangle$ and read off the anomalous dimensions as explained in section \ref{Pre}. The same information can be actually read off from the three point function $\langle \tilde J^\epsilon_s (0) \psi ^i(p)  \bar \psi^j(-p) \rangle$ by solving the RG flow equation as in \cite{Muta:1976js}, see also \cite{Giombi:2016zwa}. It is obvious to identify which Feynman diagrams for the three point function correspond to $\tilde I_7,\tilde I_8,\tilde I_9$, thus we can successfully deduce the terms proportional to $\log (x_{12}^2)$ in the integrals.
However, in this way, we loose direct connection to the bulk computation as explained in section \ref{Bulk}, which was actually the main purpose in previous works \cite{Hikida:2016wqj,Hikida:2016cla}, see also footnote \ref{muta}. Moreover, if we are interested in the overall normalization of two point function as in \cite{Diab:2016spb}, then we have to directly evaluate the complicated integrals.

We need to extend the method in \cite{Muta:1976js} to the case of the supersymmetric model deformed by \eqref{defsusy}.
As in subsection \ref{aux}, we rewrite the action by introducing auxiliary spinor fields $\eta , \bar {\eta}$ as
\begin{align}
S 
= \int d^3 x \left[ \partial_\mu \bar{\phi}^i \partial^\mu \phi_i + \bar \psi^i \slashed{\partial} \psi_i  + \bar{\eta} \mathcal{K} + \bar{\mathcal{K}}  \eta - \frac{1}{ \kappa} \bar{\eta} \eta \right]  \, .
\end{align}
The effective propagator for $\eta , \bar {\eta}$ is obtained as
\begin{align}
F_\kappa (p) = \langle \eta (p) \bar \eta (-p) \rangle _0
=  \left(- \langle \mathcal{K} (p) \bar{\mathcal{K}} (-p) \rangle_0 \right)^{-1} 
=  - \frac{16}{N} \frac{i  \slashed{p}}{|p|} \, .
\end{align}
As  in \cite{Muta:1976js} we shift the exponent as
\begin{align}
F_\kappa (p) 
=  - \frac{16}{N} \frac{ i \slashed{p}}{(p^2)^{1/2 + \Delta}} \, ,
\label{Fkappa}
\end{align}
which is the same as changing the scaling dimension of $\eta$, $\bar{\eta}$ from $3/2$ to $3/2 - \Delta$, see, e.g., \cite{Diab:2016spb}.
This also implies that interaction terms become
\begin{align}
\mu^\Delta \int d^3 x [\bar  \bar{\eta} \mathcal{K} + \bar{\mathcal{K}}  \eta ]
\label{interactions}
\end{align}
at $\kappa \to \infty$ with the renormalization scale $\mu$.

\begin{figure}
	\centering
	\includegraphics[keepaspectratio, scale=0.6]
	{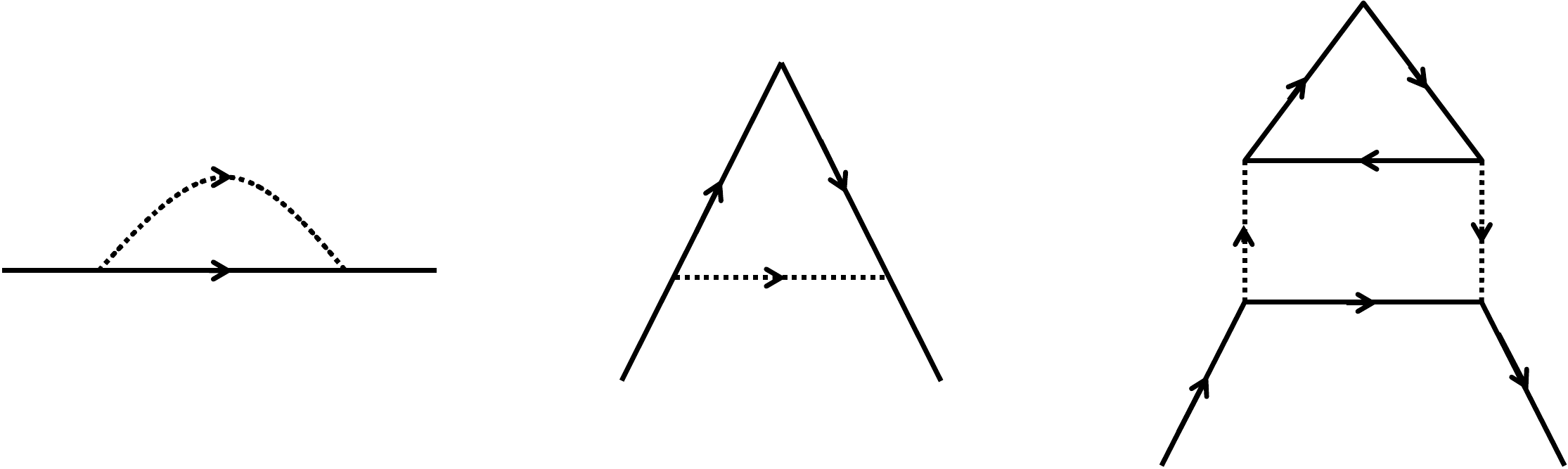}
	\caption{Feynman diagrams corresponding to $\tilde I_a$ $(a=1,\ldots,9)$.}
	\label{Muta}
\end{figure}

We read off the anomalous dimensions from the $\mu$  dependence of three point functions, such as,
$\langle \tilde J^\epsilon_s (0) \psi ^i(p)  \bar \psi^j(-p) \rangle$.
The vertex operators corresponding to $J^\epsilon_s (0) $ and $\tilde J^\epsilon_s (0) $ are
\begin{align}
V^0_{1,s} (0,p) = v_s (\hat p)^s \, , \quad
V^0_{2,s} (0,p) = \tilde v_s \hat \gamma (\hat p)^{s-1} \, , \quad
\hat p \equiv \epsilon \cdot p \, .
\label{vertices}
\end{align}
We may denote the logarithmic corrections to the vertices $V^0_{\alpha,s} (0,p)$ as
\begin{align}
\delta V^{\alpha \beta}_{s} (0,p) = - \gamma^{\alpha \beta}_s  V^0_{\beta,s} (0,p) \log \mu \, .
\end{align}
Here $\gamma^{\alpha \beta}_s$ were given in \eqref{2ptbefore}, and they become anomalous dimensions after diagonalization.%
\footnote{We should properly set the ratio $v_s/\tilde v_s$ if we want to have $\gamma^{12} = \gamma^{21}$. \label{ratio}}
There are three types of contributions as in figure \ref{Muta}. Two of them can be computed by left two Feynman diagrams, which correspond to the integrals $\tilde I_a$ $(a=1,\ldots,6)$. 
Another type of corrections come from the right most Feynman diagram.
They correspond to the integrals $\tilde I_7,\tilde I_8,\tilde I_9$, so we shall focus on the case from now on.

In the rest of this appendix, we compute the $\log \mu$ corrections $\delta V^{\alpha \beta}_s (0,p)$ corresponding to $\tilde I_7,\tilde I_8,\tilde I_9$.
Let us expand the value of Feynman integral (denoted as $X(\Delta)$) by the shift $\Delta$ introduced in \eqref{Fkappa} as 
\begin{align}
X(\Delta) = X^{(-1)} \frac{1}{\Delta} + X^{(0)}  + \mathcal{O}(\Delta) \, . 
\end{align}
We will compute the Feynman integrals of the form in the right most diagram of figure \ref{Muta}, which has $\mu^{4 \Delta}$ factor from the interaction terms \eqref{interactions}. Therefore, we can read off the $\log \mu$ dependence from the $1/\Delta$-pole term $X^{(-1)}$ as 
\begin{align}
\mu^{4 \Delta}X(\Delta) = X^{(-1)} \frac{1}{\Delta} + 4  X^{(-1)} \log \mu +  X^{(0)}  + \mathcal{O}(\Delta) \, .
\end{align}
At $\kappa \to \infty$, the contributions to anomalous dimensions can be identified with some factors in front of $\log (x_{12}^2)$, see section \ref{Pre}. Carefully treating the $(16/N)^4$ factor as in \eqref{jjf} and the normalization of higher spin currents, we would obtain
\begin{align}
& \left( \frac{16}{N} \right)^4 \tilde I_{7}^{\log} =  
\frac{16}{\pi^2 N} \frac{s}{(2s+1)(2s-1)} D_0^s \log (x_{12}^2) \, , \label{I7log} \\
& \left( \frac{16}{N} \right)^4 \tilde I_{8}^{\log} =  
\frac{16}{\pi^2 N} \frac{s}{(2s+1)(2s-1)} \tilde D_0^s \log (x_{12}^2) \, , \label{I8log} \\
& \left( \frac{16}{N} \right)^4 \tilde I_{9}^{\log} =  
\frac{16}{\pi^2 N} \frac{1}{(2s+1)(2s-1)} \frac{2 D_0^s}{s} \log (x_{12}^2)  \, . \label{I9log}
\end{align}
Here we have used the results in \eqref{dV11}, \eqref{dV22}, \eqref{dV12}, and \eqref{dV21} below.

In order to evaluate integrals, we utilize several formulas;
We introduce Feynman parameters as
\begin{align}
\frac{1}{A_1 ^{m_1} A_2^{m_2}} 
 = \int_0^1 dx \frac{(1 - x)^{m_1 - 1} x ^{m_2 - 1}}{((1-x) A_1 + x A_2)^{m_1 + m_2}} 
 \frac{\Gamma (m_1 + m_2)}{\Gamma (m_1) \Gamma(m_2)} \, .
 \label{Feynman}
\end{align}
Moreover, we use the following integrals
\begin{align}
\nonumber
&\int \frac{d^3 l}{(2 \pi)^3} \frac{1}{(l^2 + \Lambda)^n}
 = \frac{1}{8 \pi^{3/2}} \frac{\Gamma (n - 3/2)}{\Gamma (n)} 
 \left( \frac{1}{\Lambda}\right)^{n - 3/2} \, , \\
& \int \frac{d^3 l}{(2 \pi)^3} \frac{l^2}{(l^2 + \Lambda)^n}
 = \frac{1}{8 \pi^{3/2}}\frac{3}{2} \frac{\Gamma (n - 5/2)}{\Gamma (n)} 
 \left( \frac{1}{\Lambda}\right)^{n - 5/2} \, .
 \label{integrals}
\end{align}

\subsection{Correction $\delta V^{11}_s$}

Let us start from $\delta V^{11}_s$ corresponding to the integral $\tilde I_{7}$ in \eqref{tI7}.
The integral we need to evaluate is 
\begin{align}
 %\delta V^{11}_s (0,p)=
 - N  \int \frac{d^3 k}{(2 \pi)^3} \text{tr} \left[  F_\kappa (-k) A^{11}_s (k)  F_\kappa (-k) \frac{- i (\slashed{p} - \slashed{k})}{|p - k|^2} \right] \, ,
\label{V11}
\end{align}
where $F_\kappa (k)$ is given in \eqref{Fkappa} and
\begin{align}
A^{11}_s (k) =  \int \frac{d^3 l}{(2 \pi)^3} \frac{i ( \slashed{k} - \slashed{l})}{|k - l|^2} v_s (\hat l )^s \frac{1}{|l|^2} \frac{1}{|l|^2}
= i  v_s  \int \frac{d^3 l}{(2 \pi)^3} \frac{ ( \slashed{k} - \slashed{l} ) (\hat l )^s}{|l|^4 |l-k|^2} \, . \label{A11}
\end{align}

Let us work on $A^{11}_s(k)$ first.
Introducing the Feynman parameter as in \eqref{Feynman},
integration over $l$ can be performed as
\begin{align}
\nonumber
A^{11}_s (k) &=  i  v_s  \int \frac{d^3 l}{(2 \pi)^3} \int_0^1 dx \frac{\Gamma (3)}{\Gamma(1) \Gamma(2)} 
\frac{ ( 1 - x )( \slashed{k} -  \slashed{l} ) (\hat l )^s}{((l - x k)^2 + x(1-x)k^2)^3 } \\
& = 2  i  v_s  \int \frac{d^3 l ' }{(2 \pi)^3} \int_0^1 dx 
\frac{ (1 - x ) (  (1 - x ) \slashed{k} - \slashed{l} ' ) (\hat l ' + x \hat k)^s}{((l ' )^2 + x(1-x)k^2)^3 } \, .
\end{align}
Notice that the integral vanishes when the number of $l'$ in the numerator is odd and it is proportional to the metric $g^{\mu \nu}$ when the number is even. In particular, the integral vanishes when $(\hat l' )^2$ is in the numerator due to $\epsilon \cdot \epsilon = 0$.  Thus we keep $(l')^0$ and $(l')^2$ terms  as
\begin{align}
\nonumber
 (  (1 - x ) \slashed{k} - \slashed{l} ' ) (\hat l ' + x \hat k)^s
&  \sim (1-x ) \slashed{k} (x \hat k)^s - \slashed{l}'  \hat l '  s (x \hat k)^{s-1} \\
  &\sim (1-x ) \slashed{k} (x \hat k)^s - \frac13 \hat \gamma (l' )^2 s (x \hat k)^{s-1}  \, .
\end{align}
Here we have replaced $l^\mu l^\nu$ by $ l^2 g^{\mu \nu}/3$, which is possible in the numerator of the integrand.
Using \eqref{integrals}, we find
\begin{align}
A^{11}_s (k)
&=  i  \frac{1}{8 \pi^{3/2}} v_s  \int_0^1 dx 
\left[ \Gamma(\tfrac32)\frac{ ( 1 - x)^2 x^{s} \slashed{k} (\hat k)^s }{(x(1-x) k^2)^{3/2}}   - \frac{3}{2} \Gamma(\tfrac12) \frac{\frac{1}{3}  ( 1 - x ) x^{s-1} s  \hat \gamma (\hat k)^{s-1} }{(x(1-x) k^2)^{1/2}} \right] \nonumber \\
&=  i \frac{1}{16 \pi} v_s   \frac{\Gamma(\frac32) \Gamma(s - \frac12)}{\Gamma(s+1)}
\left[   \frac{\slashed{k} (\hat k)^s }{|k|^3} - s  \frac{\hat \gamma  (\hat k)^{s-1} }{ | k |} \right] \, . \label{A11result}
\end{align}

Putting \eqref{A11result} into \eqref{V11}, there are two terms in the integral over $k$.
One of them is proportional to
\begin{align}
& \int \frac{d ^3 k }{(2 \pi)^3} \text{tr} \left[ \frac{\slashed{k}}{|k|^{1 + 2 \Delta}}  \frac{\slashed{k} (\hat k)^s }{|k|^3} \frac{\slashed{k}}{|k|^{1 + 2 \Delta}}  \frac{(\slashed{p} - \slashed{k})}{|p-k|^2} \right]
 = \int \frac{d^3 k}{(2 \pi)^3} \frac{(\hat k)^s   \text{tr} [ \slashed{k} (\slashed{p} - \slashed{k}) ] }{|k|^{3 + 4 \Delta} |p-k|^2} \nonumber \\
&  \qquad \qquad =
\int \frac{d^3 k}{(2 \pi)^3} \int_0^1 dx \frac{\Gamma(\frac{5}{2} + 2 \Delta)}{\Gamma (\frac{3}{2} + 2 \Delta)} \frac{(1-x)^{1/2 + 2\Delta} (\hat k )^s \text{tr} [ \slashed{k} (\slashed{p} - \slashed{k}) ] }{((k - x p)^2 + x(1-x) p^2)^{5/2 + 2 \Delta}} \, .
\end{align}
We change $k' = k + xp$ and integrate over $k'$. The terms contributing to the $1/\Delta$-pole are
\begin{align}
 \nonumber
 (\hat k' + x \hat p)^s \text{tr} [( \slashed{k} ' + x \slashed{p} )  ((1 - x) \slashed{p} - \slashed{k} ') ]
& \sim 2s (1-2x) {\hat k} ' (x \hat p)^{s-1}  k' \cdot p  - 2 (x \hat p)^s |k'|^2 \\
& \sim \frac{2}{3}   (s x^{s-1} - (2s + 3) x^s ) ( \hat p )^s |k'|^2 \, . 
\end{align}
Thus the  $1/\Delta$-pole term is
\begin{align}
\frac{1}{2 \Delta} \frac{1}{4 \pi^2}  (\hat p )^s \left[s \frac{\Gamma (\frac32) \Gamma(s)}{\Gamma(s + \frac32)} -(2 s + 3 ) \frac{\Gamma (\frac32) \Gamma(s+1)}{\Gamma(s + \frac52)}  \right] = - 
\frac{1}{2 \Delta} \frac{1}{4 \pi^2} s (\hat p )^s \frac{\Gamma (\frac32) \Gamma(s)}{\Gamma(s + \frac32)}   \, .
\end{align}
The other integral is 
\begin{align}
\int \frac{d ^3 k }{(2 \pi)^3} \text{tr} \left[ \frac{\slashed{k}}{|k|^{1 + 2 \Delta}}  \frac{\hat \gamma (\hat k)^{s-1} }{|k|} \frac{\slashed{k}}{|k|^{1 + 2 \Delta}}  \frac{(\slashed{p} - \slashed{k})}{|p-k|^2} \right]
= \int \frac{d^3 k}{(2 \pi)^3} \frac{(\hat k)^{s-1}   ( - 2 |k|^2 (\hat p + \hat k) + 4 \hat k k \cdot p)  }{|k|^{3 + 4 \Delta}|p-k|^2}
\end{align}
up to overall factor. The first term becomes
\begin{align}
&-2 \int \frac{d^3 k}{(2 \pi)^3} \frac{(\hat k)^{s-1}   (\hat p + \hat k)  }{|k|^{1 + 4 \Delta}|p-k|^2} \nonumber \\
%  & \qquad = - 2 \int \frac{d^3 k}{(2 \pi)^3} \int_0^1 dx \frac{\Gamma (\frac{3}{2} + 2\Delta)}{\Gamma(\frac12 + 2 \Delta)}
%\frac{(1-x)^{-1/2 + 2 \Delta}(\hat k)^{s-1}   (\hat p + \hat k)  }{((k - x p)^2 + x(1-x)p^2)^{3/2 + 2 \Delta}} \\
  & \qquad = - 2 \int \frac{d^3 k ' }{(2 \pi)^3} \int_0^1 dx \frac{\Gamma (\frac{3}{2} + 2\Delta)}{\Gamma(\frac12 + 2 \Delta)}
  \frac{(1-x)^{-1/2 + 2 \Delta}(\hat k' + x \hat p)^{s-1}   ((1+x) \hat p + \hat k ')  }{((k ' )^2 + x(1-x)p^2)^{3/2 + 2 \Delta}} \, . 
\end{align}
The $1/\Delta$-pole term is then
\begin{align}
- \frac{1}{2 \Delta} \frac{1}{4 \pi^2} (\hat p)^s \left[  \frac{\Gamma (s) \Gamma (\frac12)}{\Gamma(s+\frac12)} + \frac{\Gamma(s+1) \Gamma(\frac12)}{\Gamma(s+\frac32)}  \right] \, .
\label{twoterms}
\end{align}
The second term is 
\begin{align}
& 4  \int \frac{d^3 k}{(2 \pi)^3} \frac{(\hat k)^{s} k \cdot p  }{|k|^{3 + 4 \Delta}|p-k|^2} \nonumber \\
& \qquad = 4  \int \frac{d^3 k ' }{(2 \pi)^3} \int_0^1 dx  \frac{\Gamma(\frac52 + 2 \Delta)}{\Gamma (\frac32 + 2 \Delta)} \frac{(1-x)^{1/2 + 2 \Delta}(\hat k ' + x \hat p)^{s} (k ' + x p) \cdot p  }{((k ')^2 + x(1-x)p^2)^{5/2 + 2 \Delta}} \, . 
\end{align}
The singular term arises as
\begin{align}
\frac{1}{2 \Delta} \frac{1}{2 \pi^2} s \hat p^s \frac{\Gamma(\frac32) \Gamma (s)}{\Gamma(s+\frac32)} \, ,
\end{align}
which cancels with the second term in \eqref{twoterms}.
Summing over all contributions, we find
\begin{align}
\delta V^{11}_s (0,p) = \frac{16}{\pi^2 N } \frac{s}{(2s+1)(2s-1)} V^0_{1,s} (0,p) \log \mu\, .
\label{dV11}
\end{align}

\subsection{Correction $\delta V^{22}_s$}

We move to $\delta V^{22}_s$ corresponding to the integral $\tilde I_{8}$ in \eqref{tI8}.
We need to compute the Feynman integral 
\begin{align}
 N  \int \frac{d^3 k}{(2 \pi)^3}  F_\kappa (k)  A^{22}_s (k)F_\kappa (k)  \frac{1}{|p - k|^2}  \, ,
\end{align}
where
\begin{align}
A^{22}_s (k) =  \int \frac{d^3 l}{(2 \pi)^3} \frac{i \slashed{l}}{|l|^2} \tilde v_s \hat \gamma (\hat l )^{s-1} \frac{i \slashed{l}}{|l|^2} \frac{1}{|k-l|^2}
= - \tilde  v_s  \int \frac{d^3 l}{(2 \pi)^3} \frac{ \slashed{l} \hat \gamma \slashed{l} (\hat l)^{s-1}}{|l|^4 |l-k|^2} \, .
\end{align}

We first evaluate $A^{22}_s(k)$, which can be rewritten as
\begin{align}
A^{22}_s (k) = \tilde  v_s  \int \frac{d^3 l}{(2 \pi)^3} \left[ \frac{ \hat \gamma (\hat l)^{s-1}}{|l|^2 |l-k|^2} -2 \frac{ \slashed{l} (\hat l)^s}{|l|^4 |l-k|^2} \right] \, .
\label{A22}
\end{align}
The first term becomes
\begin{align}
\nonumber
  \tilde  v_s  \int \frac{d^3 l '}{(2 \pi)^3}  \int_0^1 dx 
  \frac{ \hat \gamma (\hat l ' + x \hat k)^{s-1}}{((l ')^2 + x(1-x)k^2)^2} 
 & =   \tilde  v_s  \frac{\Gamma (\frac12)}{8 \pi^{3/2}}  \int_0^1 dx 
  \frac{ \hat \gamma ( x \hat k)^{s-1}}{(x(1-x)k^2)^{1/2}}  \\
  &=  \tilde  v_s  \frac{1}{8 \pi} \frac{\Gamma (s- \frac12) \Gamma (\frac12)}{\Gamma(s)}\frac{\hat \gamma (\hat k)^{s-1}}{|k|} \, ,
\end{align}
and the second term is
\begin{align}
\nonumber
 &- 2 \tilde  v_s  \int \frac{d^3 l '}{(2 \pi)^3}  \int_0^1 dx \Gamma (3) \frac{ (1-x) (\slashed{l'} + x \slashed{k} )(\hat l' + x \hat k )^s}{((l')^2 + x(1-x) k^2)^3} \\
& \sim - 4 \tilde  v_s  \int \frac{d^3 l '}{(2 \pi)^3}  \int_0^1 dx \frac{ (1-x) (x^{s+1} \slashed{k} (\hat k)^s + \frac13 s x^{s-1} \hat \gamma (\hat k)^{s-1}  (l')^2)}{((l')^2 + x(1-x) k^2)^3} \\
& = - 4 \tilde  v_s \frac{1}{8 \pi^{3/2}}  \int_0^1 dx \left[ \frac{ (1-x)x^{s+1} \slashed{k} (\hat k)^s}{( x(1-x) k^2)^{3/2}} \frac{\Gamma(\frac32)}{\Gamma(3)} + \frac{\frac13 s (1-x) x^{s-1} \hat \gamma (\hat k)^{s-1}  }{ (x(1-x) k^2)^{1/2}} \frac{3}{2} \frac{\Gamma(\frac12)}{\Gamma(3)}\right] \nonumber \\
&= - \tilde  v_s \frac{1}{8 \pi} \left[ \frac{ \slashed{k} (\hat k)^s}{|k|^{3}} \frac{\Gamma(\frac12)\Gamma(s+\frac12)}{\Gamma(s+1)} + \frac{1}{2} \frac{ \hat \gamma (\hat k)^{s-1}  }{ |k|} \frac{\Gamma(\frac12)\Gamma(s-\frac12)}{\Gamma(s)} \right]\nonumber \, .
\end{align}
Collecting the two contributions, we find
\begin{align}
A^{22}_s (k) =  - \tilde  v_s \frac{1}{8 \pi} \left[ \frac{ \slashed{k} (\hat k)^s}{|k|^{3}} \frac{\Gamma(\frac12)\Gamma(s+\frac12)}{\Gamma(s+1)} - \frac{1}{2} \frac{ \hat \gamma (\hat k)^{s-1}  }{ |k|} \frac{\Gamma(\frac12)\Gamma(s-\frac12)}{\Gamma(s)} \right] \, .
\end{align}

There are two terms in $A^{22}_s(k)$, and we start from the contribution to $\delta V_s^{22}$ from the first term. We compute
\begin{align}
&\int \frac{d^3 k}{(2 \pi)^3} \frac{\slashed{k}}{|k| ^{1 + 2 \Delta }}\frac{ \slashed{k} (\hat k)^s}{|k|^{3}}\frac{\slashed{k}}{|k|^{1 + 2 \Delta} } \frac{1}{|p-k|^2} 
= \int \frac{d^3 k}{(2 \pi)^3} \frac{\slashed{k} (\hat k)^s}{|k|^{3 + 4 \Delta }|p-k|^2} \nonumber  \\
&= \frac{\Gamma(\frac52 + 2 \Delta)}{\Gamma (\frac32 + 2 \Delta)} 
\int_0^1 dx \int \frac{d^3 k' }{(2 \pi)^3} \frac{(1-x)^{1/2 + 2 \Delta } (\slashed{k}' + x \slashed{p})(\hat k' + x \hat p)^s}{((k')^2 + x(1-x)p^2)^{5/2 + 2 \Delta}} \label{v22_1} \\
&\sim \frac{\Gamma(\frac52 + 2 \Delta)}{\Gamma (\frac32 + 2 \Delta)} 
\int_0^1 dx \int \frac{d^3 k' }{(2 \pi)^3} \frac{(1-x)^{1/2 + 2 \Delta } \slashed{k}' s \hat k' (x\hat p)^{s-1}}{((k')^2 + x(1-x)p^2)^{5/2 + 2 \Delta}}
= \frac{1}{2 \Delta} \frac{1}{8 \pi^2} \frac{\Gamma(\frac32) \Gamma(s+1)}{\Gamma(s+\frac32)} \hat \gamma (\hat p)^{s-1} \, .\nonumber 
\end{align}
For the contribution from the second term, we find
\begin{align}
\nonumber
& \int \frac{d^3 k}{(2 \pi)^3} \frac{\slashed{k}}{|k|^{1 + 2 \Delta} } \frac{ \hat \gamma (\hat k)^{s-1}  }{ |k|} \frac{\slashed{k}}{|k|^{1 + 2 \Delta} } \frac{1}{|p-k|^2}  \\
& \qquad =  - \int \frac{d^3 k}{(2 \pi)^3} \left[ \frac{\hat \gamma (\hat k)^{s-1}}{|k|^{1 + 4 \Delta} |p-k|^2} - 2 \frac{\slashed{k} (\hat k)^{s}}{|k|^{3 + 4 \Delta} |p-k|^2}\right] \, .
\end{align}
The second term is twice of \eqref{v22_1} and the first term becomes
\begin{align}
 &- \int \frac{d^3 k}{(2 \pi)^3}  \frac{\hat \gamma (\hat k)^{s-1}}{|k|^{1 + 4 \Delta} |p-k|^2} = - \hat \gamma \frac{\Gamma (\frac{3}{2} + 2 \Delta)}{\Gamma(\frac{1}{2} + 2 \Delta)} \int_0^1 dx \int \frac{d ^3 k'}{(2 \pi)^3} \frac{(1-x)^{-1/2 + 2\Delta} (\hat k' + x \hat p)^{s-1}}{((k')^2 + x(1-x)p^2)^{3/2 + 2 \Delta}} \nonumber \\
 &= - \frac{1}{2 \Delta} \frac{1}{8 \pi^2} \frac{\Gamma(\frac12) \Gamma (s)}{\Gamma(s+\frac12)} \hat \gamma (\hat p)^{s-1} \, .
\end{align}
Thus we find
\begin{align}
\delta V^{22}_s (0,p) = \frac{16}{\pi^2 N } \frac{s}{(2s+1)(2s-1)}  V^0_{2,s} (0,p) \log \mu \, .
\label{dV22}
\end{align}

\subsection{Corrections $\delta V^{12}_s$ and $\delta V^{21}_s$}

Finally we evaluate $\delta V^{12}_s$ and $\delta V^{21}_s$ corresponding to the integral $\tilde I_{9}$ in \eqref{tI9}.
The integrals are
\begin{align}
&  N  \int \frac{d^3 k}{(2 \pi)^3}  F_\kappa (k) A^{11}_s (k)  F_\kappa (k)  \frac{1}{|p - k|^2}  \, , \\
& - N  \int \frac{d^3 k}{(2 \pi)^3}  \text{tr} \left[ F_\kappa (- k)   A^{22}_s (k) F_\kappa (- k)  \frac{-i (\slashed{p} - \slashed{k})}{|p - k|^2} \right]  \, ,
\end{align}
respectively.
Here $A^{11}_s (k)$ and $A^{22}_s (k) $ were defined in \eqref{A11} and \eqref{A22}.  The $1/\Delta$-pole structures can be read off from the results in the previous subsections. 
The final expressions are 
\begin{align}
&\delta V^{12}_s (0,p)=  
\frac{16}{N \pi^2} \frac{1}{(2s -1 ) (2s+1)} ( \tfrac{i}{2} v_s \hat \gamma (\hat p)^{s-1}) \log \mu \, , \label{dV12} \\
&\delta V^{21}_s (0,p)=  
\frac{16}{N \pi^2} \frac{1}{(2s -1 ) (2s+1)} (- 2 i \tilde v_s (\hat p)^{s}) \log \mu \, . \label{dV21}
\end{align}
As mentioned in footnote \ref{ratio}, we need to change the relative normalization of vertices $V_{1,s}^0 (0,p)$ and $V_{2,s}^0 (0,p)$ in \eqref{vertices} so as to obtain $\gamma_{12} = \gamma_{21}$.

%\bibliographystyle{JHEP}
%\bibliography{AdS4}

\providecommand{\href}[2]{#2}\begingroup\raggedright\endgroup

\end{document}